\documentclass[12pt,preprint]{aastex}
\usepackage{graphicx}
\usepackage[usenames,dvips]{color}
\shorttitle{Cataclysmic variables above the period gap}
\shortauthors{B. Willems et al.}

\begin{document}

\title{Theoretical orbital period distributions of cataclysmic
  variables above the period gap: effects of circumbinary disks}

\author{Bart Willems}
\affil{Northwestern University, Department of Physics and Astronomy, 2131
  Tech Drive, Evanston, IL 60208, USA}
\email{b-willems@northwestern.edu}

\author{Ronald E. Taam}
\affil{Northwestern University, Department of Physics and Astronomy,
  2131 Tech Drive, Evanston, IL 60208, USA \and ASIAA/National Tsing
  Hua University - TIARA, Hsinchu, Taiwan}
\email{r-taam@northwestern.edu}

\author{Ulrich Kolb} 
\affil{The Open University, Department of Physics and Astronomy, 
  Walton Hall, Milton Keynes, MK7 6AA, UK}
\email{u.c.kolb@open.ac.uk}

\author{Guillaume Dubus}
\affil{Ecole Polytechnique, Laboratoire Leprince-Ringuet, 91128
  Palaiseau, France \and Institut d'Astrophysique de Paris, 98bis
  Boulevard Arago, 75013 Paris, France}
\email{gd@poly.in2p3.fr}

\and

\author{Eric L. Sandquist} 
\affil{San Diego State University, Department of Astronomy, 5500
  Campanile Drive, San Diego, CA 92182, USA}
\email{erics@mintaka.sdsu.edu}

\begin{abstract}
Population synthesis tools are used to investigate the population of non 
magnetic cataclysmic variable binary systems with unevolved main 
sequence-like donors for systems above the upper edge of the period gap 
at orbital periods greater than 2.75\,hr. In addition to the angular 
momentum losses associated with gravitational radiation, magnetic braking, 
and mass loss from the system, we also include the effect of a circumbinary 
disk on the evolution of the binary system.  To calculate the orbital 
period distribution, a grid of detailed binary evolutionary sequences 
has been included in the simulations. For a fractional mass input rate 
into the disk, corresponding to $3 \times 10^{-4}$ of the mass 
transfer rate, the model systems exhibit a bounce at orbital periods 
greater than 2.75\,hr.  The simulations reveal that (1) some systems can 
exist as dwarf nova type systems throughout their lifetime, (2) dwarf nova 
type systems can evolve into nova-like systems as their mass transfer rate 
increases with increasing circumbinary disk mass, and (3) nova-like systems 
can evolve back into dwarf nova systems during their postbounce evolution to longer 
orbital periods. Among these subclasses, nova-like cataclysmic variables 
would be the best candidates to search for circumbinary disks at wavelengths 
$\ga 10\,\mu$m.

The theoretical orbital period distribution of our population synthesis 
model is in reasonable accord with the combined population of dwarf novae 
and nova-like systems above the period gap, suggesting the possibility that 
systems with unevolved donors need not detach and evolve below the period gap as in 
the disrupted magnetic braking model. The resulting population furthermore 
reveals the possible presence of systems with small mass ratios (a property 
of systems exhibiting superhump phenomena at long orbital periods) and a 
preference of O/Ne/Mg white dwarfs in dwarf nova systems in comparison to nova-like 
systems.  The nova-like population furthermore shows a lack of systems with high-mass 
white dwarfs.  
The importance of observational bias in accounting for the 
differing populations is examined, and it is shown that an understanding 
of these effects is necessary in order to confront the theoretical 
distributions with the observed ones in a meaningful manner.
\end{abstract}


\section{Introduction}

Cataclysmic variable (CV) systems are a class of compact binaries in 
which a main sequence-like donor star transfers mass via Roche lobe
overflow to its white dwarf (WD) companion.  A theoretical 
understanding of their observed orbital period distribution has long been 
sought since it offers the possibility of probing the formation and 
evolution of such systems. Key features in the observed distribution 
include a dearth of systems between $\sim$ 2.2\,hr and $\sim$ 2.8\,hr 
(known as the period gap) and the existence of a critical 
period at 77\,minutes below which no systems are observed (also known as 
the period minimum).  Furthermore, the number of observed systems above 
the period gap are found to be comparable to those below the gap. 

The majority of mass losing donor stars in these systems are 
unevolved and, hence, stellar expansion resulting from nuclear 
evolution is unimportant for the binary evolution of these CVs.  
As a result, an understanding of the observed orbital period distribution 
indirectly provides constraints on the rate of orbital angular 
momentum loss necessary to promote the mass transfer rate.  In 
this regard, Shafter (1992) has suggested that since the observed CV period gap is only clearly defined for dwarf novae (DNe), an investigation 
of the period distribution of DNe rather than the 
overall CV period distribution should be used to constrain theories 
of mass transfer.  In this picture, the outburst phenomena interpreted in terms of thermal instabilities in the accretion disk, can be used to provide constraints on the mass transfer rate as a function of orbital period.  Alternatively, Martin \& Tout (2005) suggest that, in the context of the hibernation picture, the ratio of nova-like (NL) systems to DNe as a function of orbital period would provide insight into the relative time spent by a given system in such phases  (Shara et al. 1986). These relative times in turn constrain the orbital angular momentum loss rate driving the mass transfer in DNe and NLs. 

Observationally, it has been inferred that the mass transfer rates 
are correlated with the orbital period with CVs at longer orbital 
periods characterized by higher rates (e.g., Patterson 1984).  As 
a result, it is generally believed that systems above the period gap 
are driven primarily by the angular momentum losses associated with 
a magnetically coupled stellar wind whereas systems below the gap are 
primarily driven by the emission of gravitational wave radiation.  
Thus, understanding the angular momentum loss mechanism is essential 
for a proper interpretation of the observed period distribution in 
terms of CV formation and evolution.
  
Pioneering work on the application of the rotational evolution of 
slowly rotating G type stars (see, Skumanich 1972) to angular momentum 
losses in CVs was first carried out by Verbunt \& Zwaan (1981). However, 
it has become increasingly recognized that such a prescription, 
extrapolated to higher rotation rates, overestimates the angular momentum 
loss rate of single stars based on both observational (Andronov, 
Pinsonneault, \& Sills 2003) and theoretical grounds (see, e.g., Ivanova 
\& Taam 2003). On the other hand, mass transfer rates promoted by 
lower angular momentum loss rates are in conflict with the conventional 
model for the period gap in terms of a discontinuous change in magnetic 
braking when the donor star becomes fully convective (Spruit \& Ritter 
1983; Rappaport, Verbunt, \& Joss 1983) since a much too narrow 
gap would be produced (e.g., McDermott \& Taam 1989).  Furthermore, the viability 
of this picture has recently been subject to additional uncertainty 
related to the lack of discontinuous change in magnetic activity in 
low mass stars (see e.g., Donati et al. 2006). In this case of continuous magnetic braking, no gap would be formed, in sharp conflict with the observed orbital period distribution.

In view of this unsatisfactory state, other orbital angular momentum loss 
mechanisms have been explored for CV evolution, including consequential 
angular momentum losses associated with mass transfer (Barker \& Kolb 
2003) and a circumbinary (CB) disk (Spruit \& Taam 2001).  With respect 
to the latter angular momentum loss mechanism, Willems et al. (2005, 
hereafter referred to as Paper I) have recently shown that the effect 
of a CB disk can enhance the angular momentum loss rate sufficiently to 
increase the theoretical period minimum from 68 minutes for gravitational 
radiation dominated evolution to the observed value of 77 minutes.  In addition, it 
provides a natural explanation to smear out the number of 
systems at short orbital periods, thereby avoiding an accumulation of systems at 
the period minimum that plagued previous theoretical models. We note 
that the recent detection of excess infrared emission from the polars 
EF Eri, V347 Pav, GG Leo, and RX J0154 (Howell et al. 2006) 
provides observational support for the presence of cool gas (and possibly 
a CB disk) surrounding CVs. 

In this paper, we further examine the effect of CB disks on the CV 
population by carrying out a theoretical study of systems with 
orbital periods longer than 2.75\,hr. This investigation extends 
earlier work by incorporating evolutionary sequences of CVs with CB 
disks guided by calculations reported in Taam, Sandquist, \& Dubus (2003) for 
population synthesis simulations.  Thus, this study supplements our recent  
work on short-period CVs presented in Paper~I.  We focus on the non magnetic
CV population since the evolution of magnetic CVs likely involves 
processes that we leave out of consideration in our calculations such as the interaction 
of the magnetic WD with the magnetosphere of the donor star. This interaction  
can affect the rate of orbital angular momentum loss driving the mass transfer (e.g., 
Li, Wu, \& Wickramasinghe 1994), and may be responsible for the absence of a period 
gap in the subpopulation of magnetic CVs. In the next section, we describe the input 
physics and parameters underlying our population synthesis calculations.  
The observed period distribution is presented in \S 3 and our treatment 
of selection effects for treating DNe and NL are described in \S 4.  
The present-day CV birth rates and numbers of systems currently populating 
the Galaxy are discussed in \S 5. In \S 6 and \S 7, the theoretical 
orbital period distributions and variations of system properties (mass transfer 
rate, donor mass, WD mass, mass ratio) as a function of orbital period 
resulting from our calculations are presented and described.  The issue of CB 
disk detectability of these systems is considered in \S 8.  In the last section, 
finally, we discuss the implications of our results and give some 
concluding remarks. 
 
\section{Input Physics and Parameters}

As outlined in detail in Paper~I, we use a hybrid binary population
synthesis technique in which the BiSEPS binary population synthesis
code (see Willems \& Kolb 2002, 2004) is used to construct a
population of zero-age CVs, and a state-of-the art stellar and binary
evolution code is used to evolve the population up to the current
epoch. For the construction of the zero-age population, the initial
primary masses $M_1$ are assumed to be distributed according to an
initial mass function similar to that of Kroupa, Tout, \& Gilmore
(1993). The distribution of initial secondary masses $M_2$ is obtained
from a power-law distribution for the initial mass ratio $q=M_2/M_1$,
or from the same initial mass function as adopted for the primary
masses. The distribution of initial orbital separations is assumed to
be logarithmically flat. These distributions are supplemented with a
constant star formation rate normalized so that one binary with $M_1 >
0.8\,M_\odot$ is born each year. The age of the Galaxy is assumed to
be 10\,Gyr, and all stars are assumed to be formed in binaries. We also 
limit ourselves to Population~I chemical compositions.

The main evolutionary processes affecting the formation of CVs are the
common envelope (CE) phase leading to the formation of the WD and the
orbital angular momentum losses driving the post-CE binary to the CV
stage. 

As is customary, we model the CE phase by equating the binding energy
of the envelope to the energy lost from the orbit due to frictional
forces as
\begin{equation}
{{G \left( M_{\rm c} + M_{\rm e} \right) M_{\rm e}}
  \over {\lambda_{\rm CE}\, R_{\rm L}}} =
  \alpha_{\rm CE} \left[ {{G\, M_{\rm c}\, M_2} \over {2\, a_{\rm f}}}
  - {{G \left( M_{\rm c} + M_{\rm e} \right) M_2}
  \over {2\, a_{\rm i}}} \right]\!  \label{ce}
\end{equation}
(Tutukov \& Yungelson 1979, Webbink 1984). Here, $G$ is the Newtonian gravitational constant;
$M_{\rm c}$, $M_{\rm e}$, and $R_{\rm L}$ are the core mass, envelope
mass, and Roche lobe radius of the WD progenitor at the start of the
CE phase; $M_2$ is the mass of the companion star; and $a_{\rm i}$ and
$a_{\rm f}$ are the initial and final orbital separations at the start
and at the end of the CE phase, respectively. Moreover, $\lambda_{\rm
  CE}$ determines the binding energy of the envelope, and $\alpha_{\rm
  CE}$ the fraction of the orbital energy transferred to the envelope.
As in Paper~I, we investigate the sensitivity of our results to the
modeling of the CE phase by considering different prescriptions for
the determination of $\alpha_{\rm CE}$ and $\lambda_{\rm CE}$. In our
standard population synthesis model (model~A), we set $\alpha_{\rm
  CE}\, \lambda_{\rm CE}=0.5$. The other prescriptions are summarized
in Table~\ref{models} and described in detail in Paper~I.

After the CE phase, orbital angular momentum losses due to
magnetic braking and/or gravitational radiation drive the systems
towards the start of the CV phase. Since the primary aim of this paper
is to investigate the effects of orbital angular momentum losses due
to CB disks, we adopt a simplified prescription
assuming a constant magnetic braking time scale of 10\,Gyr for all
main-sequence stars with masses between $0.35\,M_\odot$ and
$1.25\,M_\odot$. For stars less massive than $0.35\,M_\odot$ or more
massive than $1.25\,M_ \odot$, magnetic braking is assumed to be
ineffective. We note that, once mass transfer starts, in the majority 
of cases the CB disk dominates the angular momentum losses by the 
point in the evolution where the donor is less massive than $0.35 M_\odot$.
To determine the sensitivity of the population synthesis to the 
magnetic braking time scale, we explore in \S 9 a simulation for which 
the time scale is reduced to $4 \times 10^9$ years.

The evolution of the binaries after the onset of the CV phase is
followed using detailed evolutionary tracks obtained from a full
binary evolution code described in Taam et al. (2003). The models 
adopted for the donor stars use solar abundances. The CV phases are 
furthermore modeled assuming fully non-conservative mass transfer as a 
result of efficient mass loss during nova explosions. The mass of the 
WD is then fixed throughout the evolution of the CVs. Based upon the 
work of Taam et al. (2003), the fractional amount of mass 
deposited in the CB disk, $\delta$, must be higher for systems above the 
period gap than below the period gap. 
The evolution of the CB disk and its implementation into the
binary evolutionary code is described in Dubus, Taam, \& Spruit (2002)
and Taam et al. (2003).  Hence, we adopt the same model as in 
Paper I except that we consider CVs with orbital periods longer than 
2.75 hr and, guided by the results of Taam et al. (2003), $\delta = 3 
\times 10^{-4}$. 

For further details on the computational technique and the adopted input
physics, we refer to Paper~I.

\section{The Observed Orbital Period Distribution Above 2.75\,hr}

\begin{figure}
\center
\resizebox{8.0cm}{!}{\includegraphics{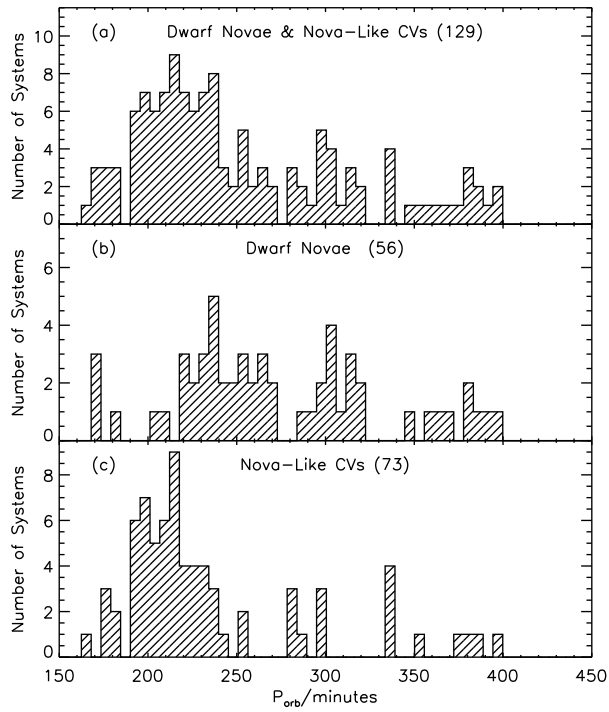}}
\caption{Observed orbital period distribution of non-magnetic CVs with
  periods between 2.75\,hr and 6.7\,hr, and their subdivision into DNe and NLs.} 
\label{obs}
\end{figure}

Before presenting the results of the theoretical population synthesis calculations, we briefly discuss the main features of the observed orbital period distribution of non-magnetic CVs with periods longer than 2.75\,hr. For this purpose, we extract all DNe and NLs from the 2006 January edition (RKCat 7.6) of the Ritter \& Kolb (2003) catalog of cataclysmic binaries, low-mass X-Ray binaries and related objects.  The resulting
orbital period distribution of non-magnetic CVs is shown in panel~(a)
of Fig.~\ref{obs}. Its subdivision into DNe and NLs is shown in
panels~(b) and~(c) respectively. In each panel the number between parentheses indicates the total number of systems in the sample.

The distribution of non-magnetic CVs decreases rapidly with decreasing orbital periods between 160 to 180\,minutes, corresponding to the upper
edge of the period gap. It also shows a broad maximum at periods of
190--240\,minutes, and a slowly decreasing tail at periods longer than 
240\,minutes. For the purpose of this paper, we restrict the observed
CV population to systems with periods of less than 400\,minutes. Systems 
with longer orbital periods typically have significantly evolved donor 
stars (see Baraffe \& Kolb 2000) which introduce additional
complications in the theoretical population synthesis calculations
(see Kolb \& Willems 2005 for some initial results on zero-age
CVs with evolved donor stars).

When DNe and NLs are considered separately, two considerably different
distributions emerge. The DN population exhibits three peaks of
decreasing height at periods of 240, 300, and 380\,minutes. The number
of systems contributing to the distribution is, however, rather 
limited. Whether these peaks are representative of the full population 
or a consequence of small number
statistics therefore remains to be confirmed by future
observations. For this reason, we will focus the population synthesis
calculations on explaining the global increase in the number of
systems with decreasing orbital periods from 400 to 240\,minutes, and
the subsequent decline in the number of systems for periods
smaller than 240\,minutes.

The NL population, on the other hand, shows a broad peak in the
orbital period distribution at periods of 190--220\,minutes. The distribution
decreases rapidly on both sides of the peak, and shows possibly a hint of a
secondary peak at 340\,minutes. The statistics of the sample near this
period are, however, again not sufficient to conclusively confirm or
refute the presence of the peak.

\section{Observational Selection Effects}
\label{selection}

The inherent differences between the populations of DNe (transient systems spending most of their life time in quiescence) and NLs (persistently bright systems)
likely subject them to considerably different observational selection
effects. In order to cope with them, we separate the CVs in the
population synthesis calculations into transient and persistent
sources according to the critical mass-transfer rate for optically
thick accretion disks derived by Hameury et al. (1998):
\begin{equation}
\dot{M}_{\rm crit} = 9.5 \times 10^{15}\, \alpha^{0.01} 
  \left( {M_{\rm WD} \over M_\odot} \right)^{-0.89} \left( 
  {r_{\rm disk} \over {10^{10}\,{\rm cm}}} \right)^{2.68} 
  {\rm g\,s^{-1}}.  \label{mdotcrit}
\end{equation}
Here, $\alpha$ is the viscosity parameter, $M_{\rm WD}$ the mass of
the WD, and $r_{\rm disk}$ the radius of the accretion disk. In our
calculations, we set $\alpha=0.01$ and $r_{\rm disk}=(2/3)R_{\rm
L,WD}$, where $R_{\rm L,WD}$ is the Roche-lobe radius of the
WD. Systems with mass-transfer rates below the resulting
$\dot{M}_{\rm crit}$ then correspond to DNe, while systems with
mass transfer rates above $\dot{M}_{\rm crit}$ correspond to 
NLs.

In Paper~I, we modeled the selection effects biasing the detection of DNe 
assuming an isotropic distribution of systems in the Galactic disk. For a 
bolometric luminosity limited sample, the total observable volume then scales 
as the accretion luminosity $L_{\rm acc}$ raised to the power 1.5. Dwarf novae 
above the period gap can, however, reach outburst luminosities high enough 
to make them visible at distances large enough for the finite scale height of 
the Galactic disk to affect their observed spatial distribution. The assumption 
of an isotropic distribution is then more appropriately replaced by an 
axisymmetric distribution limited by a fixed scale height vertical to the
Galactic plane, so that the total observable volume for a bolometric
luminosity limited sample is linearly proportional to $L_{\rm acc}$. For simplicity, we 
consider both possible observational selection factors and compare the 
theoretical orbital period distributions obtained by weighting the 
contribution of each system to the population according to $L_{\rm acc}^{1.5}$ 
with the distributions obtained by weighting the contribution of each system 
according to $L_{\rm acc}$.

The DN accretion luminosity is determined as
\begin{equation}
L_{\rm acc} = {{G\,M_{\rm WD}\,\dot{M}_{\rm out}} \over {R_{\rm WD}}}, \label{d1}
\end{equation}
where $\dot{M}_{\rm out}$ is the mass accretion rate during outbursts, and $R_{\rm WD}$ is the radius of the WD. We approximate $\dot{M}_{\rm out}$ as
\begin{equation}
 \dot{M}_{\rm out} = \dot{M}_d/d,  \label{d2}
\end{equation} 
where $\dot{M}_d$ is the secular mean mass-transfer rate, and $d=t_{\rm out}/t_{\rm rec}$ is the duty cycle determined as the ratio of the duration ($t_{\rm out}$) and the recurrence time ($t_{\rm rec}$) of the outbursts. Since the discovery of DNe likely depends on the duration and recurrence time of the outbursts in addition to the magnitude of the accretion luminosity, we also weight the contribution of each system to the population according to its duty cycle $d$. The total weighting factor applied to each system is thus given by $W=d\,L_{\rm acc}$ or $W=d\,L_{\rm acc}^{1.5}$. For simplicity, we adopt a constant duty cycle, $d=0.1$, for the majority of the population synthesis models. 
To determine the sensitivity of this choice we also consider a limited number of models with 
variable duty cycle $d=\dot{M}_d/\dot{M}_{\rm crit}$.  In the case where $W = d\,L_{\rm acc}$ 
the weighting factor in these models reduces to $W = GM_{\rm WD}\,\dot{M}_d/R_{\rm WD}$ 
which is independent of $d$ and varies continuously from transient to persistent systems 
(see below). 

As an alternative weighting factor, we also explore the effect of replacing the accretion luminosity $L_{\rm acc}$ by the absolute visual luminosity $L_{\rm vis}$ determined from the empirical $L_{\rm vis}$--$P_{\rm orb}$ relations, where $P_{\rm orb}$ is the orbital period, derived by Warner (1987) or Harrison et al. (2004). 

Nova-like CVs, on the other hand, are persistently bright and thus always tend to
be visible up to distances large enough for the finite scale height of the Galactic disk to affect their distribution. As argued above, the total observable volume for a bolometric luminosity limited sample then scales as $L_{\rm acc}$, where $L_{\rm acc}$ is now given by 
Eqs.~(\ref{d1}) and (\ref{d2}) with $d=1$. We therefore model the selection effects operating on NLs by weighting the contribution of each system to the population by a factor proportional to $L_{\rm acc} = GM_{\rm WD}\,\dot{M}_d/R_{\rm WD}$. A similar weighting factor is obtained if one adopts
a visual magnitude limited sample and uses power-law fits to the bolometric correction to relate the visual magnitude to the bolometric luminosity and the distance of the source from the Sun. (D\"{u}nhuber 1994, Kolb 1996).

\section{Present-Day Galactic CV Population}

\subsection{Birth Rates}
\label{brates}

The characteristics of the present-day birth rate of zero-age CVs as a
function of the orbital period have been discussed extensively in
Paper~I. The dependence of the birth rate on the orbital period was
found to be in good agreement with that derived by de Kool (1992) and
Politano (1996). Here, we therefore limit ourselves to the
presentation of the numbers of CVs presently forming at periods above
2.75\,hr as well as its subdivision into the numbers of CVs forming with 
He, C/O, and O/Ne/Mg WDs. These birth rates are listed in 
Table~\ref{br} which complements the present-day birth rates of CVs
forming at periods below 2.75\,hr listed in Table~2 of Paper~I. For ease of comparison, the latter birth rates are repeated in the last column of Table~\ref{br}. 

The total number of CVs forming at orbital periods longer than 2.75\,hr is of
the order of $10^{-3}\,{\rm yr^{-1}}$ for all initial mass ratio or
initial secondary mass distributions considered, except for $n(q) \propto
q^{-0.99}$. In the latter case, the present-day birth rate of Galactic
CVs is smaller by an order of magnitude because the $n(q) \propto
q^{-0.99}$ distribution favors small initial secondary masses for which
survival of the CE phase is much more difficult than for higher initial secondary masses. 

The relative contributions of systems containing He, C/O, and O/Ne/Mg
WDs to the zero-age CV population above 2.75\,hr depend strongly on the adopted CE
model and only weakly on the adopted initial mass ratio or initial
secondary mass distribution. The relative number of systems born with
He WDs and periods longer than 2.75 hr is typically of the order of 10--15\%, except for model~CE8 in
which the post-CE orbital separation of systems containing He WDs is often 
too large for them to evolve into a CV within the life time of
the Galaxy. For model~CE8, the He WD systems therefore constitute only
a few tenths of a per cent of the zero-age CV population. Systems
containing C/O WDs, on the other hand, make up 80--90\% (95--99\% in the case of population synthesis model~CE8) of the population of newborn CVs above periods of 2.75\,hr. Systems containing O/Ne/Mg WDs, finally, are generally the
least abundant group, constituting at most a few per cent of the
zero-age population. Their contribution to the population is almost completely negligible for model CE1 in which $\alpha_{\rm CE}\, \lambda_{\rm CE}$ is too small to successfully eject the WD progenitor's massive envelope from the system. 
  
Adding the birth rates listed in Table~\ref{br} to the birth rates
derived in Paper~I for CVs forming at periods below 2.75\,hr yields
total birth rates of Galactic CVs that are in excellent agreement with those derived by de Kool (1992), Politano (1994), and Hurley, Tout, \& Pols (2002).

\subsection{Population Statistics}

In Table~\ref{num}, we list the total number of DNe and NLs with
orbital periods longer than 2.75\,hr currently populating the Galaxy,
as well as their decomposition according to the type of the WD accretor. The number of DNe and NLs is typically of the order of $10^5$--$10^6$, with the DNe being a factor of 2 to 3 more abundant than NLs. The total number of non-magnetic CVs (DNe+NLs) with periods above 2.75\,hr is furthermore about an order of magnitude smaller than the total number of non-magnetic CVs with periods below 2.75\,hr ($\sim 10^6$--$10^7$, see Table~3 in Paper~I). 

In accord with the CV birth rates, the majority ($\ga$90\%) of the
present-day population of DNe and NLs consists of systems containing
C/O WDs. Systems containing He WDs typically comprise less than 10\% of 
the population due to mass transfer stability. 
The largest differences between the DN and NL populations
occur for the O/Ne/Mg WD systems which are 1 to 2 orders of magnitude
more abundant in the DN population than in the NL population. This is
directly related to two factors. First, for a given mass donor, the 
secular mean mass-transfer $\dot{M}_d$ decreases with  decreasing mass 
ratio $q=M_d/M_{\rm WD}$, where $M_d$ is the mass of the donor and 
$M_{\rm WD}$ the mass of the WD.  Secondly, the critical mass-transfer 
rate $\dot{M}_{\rm crit}$ increases rapidly with the increasing size of 
the white dwarf Roche lobe and, therefore, with decreasing $q$. It is 
also interesting to note that the contribution of He WD systems to the 
population of DNe shows a strong dependence on the
adopted CE model and only a weak dependence on the adopted initial
mass ratio or initial secondary mass distribution, while the
contribution of O/Ne/Mg WD systems shows a strong dependence on the
adopted initial mass ratio or initial secondary mass distribution but
not on the adopted CE model (except for model CE1). 

\begin{figure}
\center
\resizebox{8.0cm}{!}{\includegraphics{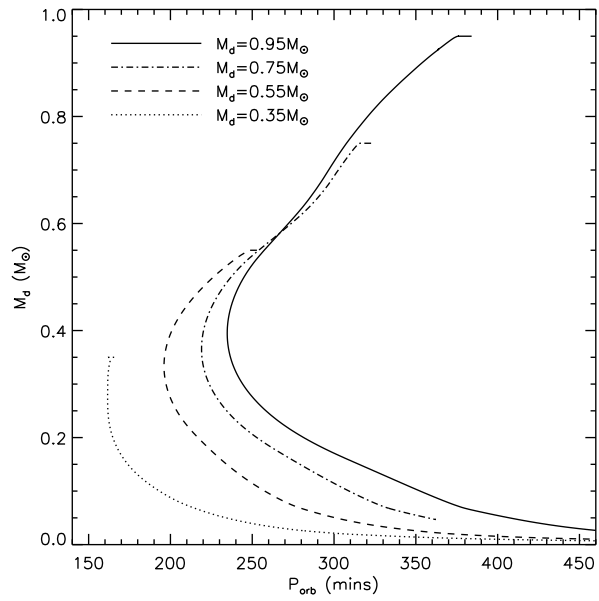}}
\caption{Donor mass as a function of orbital period for CVs evolving under the influence of CB disks with a fractional mass input rate of $3 \times 10^{-4}$. The WD mass is taken to be $0.6\,M_\odot$.} 
\label{tracks}
\end{figure}

The typical evolution of a CV surrounded by a CB disk starts off with 
an evolution from longer to shorter orbital periods driven by 
orbital angular momentum losses, followed by an evolution from shorter to 
longer orbital periods as the donor star is significantly driven out of 
thermal equilibrium by the mass loss from its surface. The transition from 
decreasing to increasing orbital periods occurs when $(\partial \ln 
R_d)/(\partial \ln M_d)=1/3$, where $M_d$ and $R_d$ are the mass and radius 
of the donor star (Taam et al. 2003).  The evolution is illustrated in 
Fig.~\ref{tracks} for CVs with $M_{\rm WD}=0.6\,M_\odot$ and a CB disk mass 
input rate equal to $3 \times 10^{-4}$ times the donor's mean secular mass transfer rate. 
The different curves correspond to donor stars with initial masses $M_d = 
0.35, 0.55, 0.75, 0.95\,M_\odot$. The ``bounce" period at which the systems 
start evolving towards longer orbital periods is seen to increase with 
increasing initial donor mass due to the dependence of the orbital angular 
momentum loss rate on the CB disk mass (which in turn depends on the initial donor mass). 
 
Hence, when CB disks are incorporated in the evolution of CVs, the population of systems above the period gap can be divided into systems evolving from long to short orbital periods (prebounce systems) and systems evolving from short to long orbital periods
(postbounce systems). The relative number of pre- and postbounce systems and their decomposition
into systems containing He, C/O, and O/Ne/Mg WDs are listed as percentages in
Table~\ref{prepost}. In general, the DN population consists of more
prebounce than postbounce systems, while the NL population consists of
more postbounce than prebounce systems. This reflects
the fact that during the early phases of mass transfer, all systems have
mass-transfer rates low enough to be transients, while during the
later phases only some have mass-transfer rates low enough to be
transients. Our models furthermore predict that almost all DNe and NLs
containing He WDs should be postbounce rather than prebounce systems. 
This follows from the narrow range of initial donor masses ($\sim 0.35\,M_\odot$ 
to $\sim 0.45\,M_\odot$) giving rise to stable mass transfer in systems containing 
He WDs above the upper edge of the period gap and the bounce of these systems 
subsequent to the onset of mass transfer (see Fig.~\ref{tracks}).

\section{Theoretical Orbital Period Distributions}

As in Paper~I, we present the results of the CV population synthesis
calculations in two steps: we first discuss the intrinsic present-day
population, and next the theoretically expected observed present-day population. 
Both steps focus on CVs forming at periods longer than 2.75\,hr and neglect 
any systems formed below 2.75\,hr that are evolving to longer orbital periods 
(based on the results of Paper I, less than 2\% of the CVs forming below 2.75\,hr 
evolve to periods longer than 2.75\,hr). For ease of comparison, all distributions 
presented in the following subsections are normalized to unity.

\subsection{The Intrinsic Orbital Period Distribution}

We first consider the present-day CV population obtained by evolving
the zero-age population up to the current epoch. At this stage, we do 
not yet account for any observational selection effects that may be affecting 
the observed orbital period distribution. The selection effects are treated 
separately and discussed in detail in \S\,\ref{obssel}.

The resulting probability distribution functions (PDFs) and cumulative
distribution functions (CDFs) of the orbital periods of the combined
population of DNe and NLs (i.e. all non-magnetic CVs) evolving under the influence of a
CB disk are shown in Fig.~\ref{pint0}. For comparison, the observed orbital period distributon is shown by means of a thick solid line. The degree of agreement or disagreement between the theoretical and observed orbital period distributions is indicated by the Kolmogorov-Smirnov (KS) significance level $\sigma_{\rm KS}$ in the CDF panels of the figure. Here $\sigma_{\rm KS}=0$ indicates large differences between the theoretical and observed distributions, while $\sigma_{\rm KS}=1$ indicates excellent agreement between the two distributions. The results for the various population synthesis models listed in Table~\ref{models} are all qualitatively similar to those of population synthesis model~A, regardless of the adopted initial mass ratio distribution. An initial secondary mass distribution equal to the initial primary mass distribution furthermore yields orbital period distributions similar to those resulting from the initial mass ratio distribution $n(q) \propto q^{-0.99}$. In this and the following sections, we therefore restrict the presentation and discussion of the theoretical orbital period distributions to the PDFs and CDFs obtained for population synthesis model~A and the initial mass ratio distributions $n(q) \propto q^{-0.99}$, $n(q)=1$, and $n(q) \propto q$. 

For a flat initial mass ratio distribution $n(q)=1$, the theoretical orbital period distribution of non-magnetic CVs increases smoothly with decreasing period from 400 to 240\,minutes, and shows a broad peak at periods of 180--240\,minutes. The same applies to the theoretical orbital period distribution obtained for $n(q) \propto q$, except that the peak is replaced by a cut-off at 160\,minutes. For $n(q) \propto q^{-0.99}$, the simulated PDF reaches a maximum at $\sim 170$\,minutes. All three theoretical distributions furthermore show a local plateau at $P_{\rm orb } \simeq 300$\,minutes. 

In accord with the relative numbers of systems discussed in the previous section, the theoretical orbital period distributions are dominated by systems containing C/O WDs. Systems containing He WD accretors provide a small contribution at periods of 180--200\,minutes. A small contribution from systems containing O/Ne/Mg WDs is also visible when $n(q) \propto q^{-0.99}$.

The PDFs and CDFs of the separate DN and NL CV populations are shown in the left- and right-hand panels of Fig.~\ref{pint}, respectively.  For $n(q)=1$ and $n(q) \propto q^{-0.99}$, the orbital period distribution of DNe displays a rapid increase in the number of systems with decreasing period from 400 to 300\,minutes, followed by a much slower increase from 300 to 160\,minutes. For $n(q) \propto q$, the simulated orbital period distribution is almost flat between 180 and 300\,minutes. The simulated orbital period distributions of Galactic NLs, on the
other hand, show a broad peak at 230\,minutes with a rapid decline
towards shorter orbital periods and a slower decline towards longer
orbital periods.  

The decomposition of the orbital period distribution of non-magnetic CVs 
into distributions for DNe and NLs depends strongly on the adopted critical 
mass-transfer rate separating transient from persistent systems. In Fig.~\ref{mdc} we therefore illustrate the effects of uncertainties in $\dot{M}_{\rm crit}$ on the theoretical DN and NL orbital period distributions by increasing and decreasing $\dot{M}_{\rm crit}$ by a factor of 3 with respect to the $\dot{M}_{\rm crit}$ given by Eq.~(\ref{mdotcrit}). The solid lines correspond to the orbital period distributions shown in Fig.~\ref{pint} for population synthesis model~A and the initial mass ratio distribution $n(q)=1$. The dash-dotted and dotted lines correspond to the orbital period distributions obtained for critical mass transfer rates that are a factor of 3 smaller and larger, respectively. 

Decreasing $\dot{M}_{\rm crit}$ by a factor of 
3 significantly flattens the intrinsic DNe orbital period distribution, 
while increasing  $\dot{M}_{\rm crit}$ by a factor of 3 broadens the 
cut-off at 160\,minutes to a peak spanning periods from 160 to 240\,
minutes. For the NL population, decreasing  $\dot{M}_{\rm crit}$ by a 
factor of 3 shifts the peak at 230\,minutes to slightly lower period. Increasing  $\dot{M}_{\rm crit}$ by a factor of 3, on 
the other hand, shifts the peak in the theoretical NL orbital period 
distribution to periods of 280--300\,minutes.

\subsection{The Orbital Period Distribution Corrected for Observational Selection Effects}
\label{obssel}

The present-day orbital period distributions of DN and NL CVs accounting
for observational selection effects are shown in Fig.~\ref{pwgt}. The DN distribution is obtained by weighting the contribution of each CV in the population according to the accretion luminosity raised to the power 1.5 and a constant duty cycle $d=0.1$ (top left panel)\footnote{Note that for normalized distributions, the actual value of a constant $d$ does not play a role.},  according to the accretion luminosity raised to the power 1.5 and the duty cycle $d=\dot{M}_d/\dot{M}_{\rm crit}$ (top right panel), or according to a weighting factor that is linearly proportional to the accretion luminosity and the duty cycle $d=\dot{M}_d/\dot{M}_{\rm crit}$ (bottom left panel). The weighting factor for CVs in the NL population (bottom right panel) is taken to be linearly proportional to the accretion luminosity. 
 
All three of the observational selection factors considered for DNe diminish the abrupt cut-off in the intrinsic PDFs at $P_{\rm orb} \simeq 160$\,minutes and significantly increase the contribution of systems containing O/Ne/Mg WDs to the theoretical orbital period distribution. For $n(q) \propto q^{-0.99}$, the selection factors introduce a local maximum at 190\,minutes which is at too short of an orbital period to explain the peak in the observed orbital period distribution at 240\,minutes, while for $n(q) \propto q$, the selection factors introduce a small peak at 300\,minutes which coincides with the secondary peak in the observed distribution. The initial mass ratio distribution $n(q)=1$, on the other hand, yields an almost flat distribution between 180 and 300\,minutes. 
Based on the KS measure, the initial mass ratio distributions $n(q)=1$ and 
$n(q) \propto q$ yield the most satisfactory agreement with the observed 
orbital period distribution.  Regardless of the adopted initial mass 
ratio distribution, the $W=d\,L_{\rm acc}=GM_{\rm WD}\,\dot{M}_d/R_{\rm WD}$  
weighting factor (bottom left panel) gives the most satisfactory agreement between the 
steepness of the theoretical and the observed orbital period distribution at 
$\sim 170$\,minutes (the upper edge of the period gap).

Adopting the empirical $L_{\rm vis}$--$P_{\rm orb}$ relations of Warner (1987) or Harrison et al. (2004) in the weighting factors instead of the accretion luminosity $L_{\rm acc}$ does not significantly affect the overall theoretical DN orbital period distributions compared to the corresponding intrinsic distributions. The main difference between using $L_{\rm vis}$ and $L_{\rm acc}$ is that the empirical $L_{\rm vis}$ relation depends solely on the orbital period $P_{\rm orb}$, so that the relation at best captures the average dependence of $\dot{M}_d$ on $P_{\rm orb}$. The accretion luminosity $L_{\rm acc}$, on the other hand, explicitly depends on the mass-transfer rate and  the mass and radius of the WD, allowing for a spread in the weighting factor at a given orbital period. As a consequence, the $L_{\rm acc}$ weighting factors favor the relative contribution of higher mass WDs to the population of DNe.

The observational selection factors adopted for the NL population do not considerably alter the shape of the orbital period distribution with respect to that of the intrinsic population. The main effects of the selection factors are a narrowing of the peak at 230\,minutes, and a decrease of the relative contribution of systems containing He WDs.  The peak at 230\,minutes also shifts to slightly longer orbital periods which is opposite to the trend required to improve the agreement between theory and observations.  The deficit of NLs at long orbital periods as compared to the observed distribution likely results from our lack of including nuclearly evolved systems in the population synthesis. As a result, the level of disagreement inferred from the KS measure should not be taken at face value 
for models in which the maximum deviation between the theoretical and 
observed orbital period distributions occurs in the long-period tail of the 
distributions. However, the relative variations of this measure can be used to differentiate between the different models. In this capacity, the KS measure  favors the initial mass 
ratio distribution $n(q) \propto q^{-0.99}$, which is opposite to the initial 
mass ratio distribution favored by the KS measure for DN type systems. 
This apparent disparity in the 
best fitting initial mass ratio distributions can possibly be attributed 
to uncertainties in the binary evolution model.

\section{Distributions of System Properties as a Function of Orbital Period}
\label{prop} 

While the orbital period is by far the most accurately determined parameter of 
observed CVs and thus best suited to compare theory to observation, it is 
instructive to also consider the distribution of other system properties such 
as the mass transfer rate and component masses as a function of the orbital 
period. As before, a comparison of theoretical with observed distributions 
requires a model for the observational selection effects. However, as the 
latter is still subject to considerable uncertainties, we here focus on the 
properties of the intrinsic theoretical distributions. These distributions 
may be compared with observation once samples complete within a given volume 
of space become available.

In Fig.~\ref{mtint}, we show the intrinsic distributions of the mass-transfer rates in the simulated DN and NL CV populations as a function of the orbital period\footnote{Note that the PDFs show some signs of the finite number ($\sim 1300$) of tracks used in the calculations, especially during the early (prebounce) stages of mass transfer. This does not affect the numbers and orbital period distributions presented in the previous sections because the initial $\left(M_1, M_2, P_{\rm orb}\right)$ parameter space is very well sampled.}, for population synthesis model~A and a flat initial mass ratio distribution $n(q)=1$. Typical mass-transfer rates are of the order of 1--$2 \times 10^{-9}\,M_\odot\,{\rm yr^{-1}}$ for DNe, and 2--$6 \times 10^{-9}\,M_\odot\,{\rm yr^{-1}}$ for NLs, which is comparable to the typical mass transfer rates found by Howell, Nelson, \& Rappaport (2001) based on the magnetic braking rate of Rappaport et al. (1983) with $\gamma=3$. 
In both cases, the rates increase with increasing orbital period, although for the DNe there is also a non-neglible number of prebounce systems for which the mass-transfer rates increase with decreasing orbital period. Accounting for observational selection effects  reduces the relative number of DNe with mass-transfer rates below $10^{-9}\,M_\odot\,{\rm yr}$, but otherwise does not significantly change the DN and NL ($\dot{M}_d$, $P_{\rm orb}$) distributions displayed in Fig.~\ref{mtint}. 

The intrinsic distributions of DN and NL CV donor masses as a function of the orbital period for population synthesis model~A and a flat initial mass ratio distribution $n(q)=1$ are shown in Fig.~\ref{mdint}. The majority of DNe are prebounce systems with donor masses above $0.3\,M_\odot$. The NL population on the other hand shows a large number of postbounce systems with donor masses below $0.3\,M_\odot$. Because of the increase of the critical mass transfer rate separating transient from persistent sources with increasing orbital period, no NLs are found with donor masses below $0.1\,M_\odot$. Due to the period bounce at the upper edge of the period gap, the shape of the distributions is also intrinsically different from the shape of the distributions obtained for CVs evolving under the influence of gravitational radiation, magnetic braking, and mass loss during nova explosions only (see Fig. 5 in Howell et al. 2001). The main effect of adopting observational selection factors is to boost the relative contribution of systems with donor masses $M_d \la 0.3\,M_\odot$ in the population of DNe. These systems are all postbounce systems and thus have high mass-transfer rates and accretion luminosities favoring their detection. 

In Fig.~\ref{mwdint}, we show the intrinsic distributions of WD masses in DN and NL CVs  as a function of the orbital period for population synthesis model~A and a flat initial mass ratio distribution $n(q)=1$. In both cases, the gap at $\sim 0.5\,M_\odot$ divides systems with He WDs ($M_{\rm WD} \la 0.5\,M_\odot$) from systems with C/O WDs ($M_{\rm WD} \ga 0.5\,M_\odot$). Both DN and NL systems are clearly dominated by systems with $\sim 0.55\,M_\odot$ C/O WDs. The NL population furthermore shows a lack of systems with high-mass WDs, which gets more pronounced with increasing orbital periods. The boundary on the right of the $M_{\rm WD}$--$P_{\rm orb}$ parameter space corresponds to the transition from persistent to transient behavior due to the increase of the critical mass transfer rate with increasing orbital period. Indeed, systems near this boundary are all post-bounce systems, so that longer orbital periods correspond to lower donor masses. Correspondingly, persistent behavior requires the WD mass to decrease with increasing orbital period.

Figure~\ref{qint}, finally, shows the intrinsic distributions of mass 
ratios $q=M_d/M_{\rm WD}$ as a function of the orbital period, for 
population synthesis model~A and a flat initial mass ratio distribution 
$n(q)=1$. Typical mass ratios near the upper edge of the period gap 
are of the order of $q=0.4$--0.7 for both DNe and NLs. 
The additional orbital angular momentum losses caused by CB disks 
and the associated bounce of systems at the upper edge of the period 
gap furthermore allow for significantly smaller mass ratios for 
systems above the period gap than the standard gravitational radiation 
and magnetic braking driven CV evolution (see Fig. 6 and 11 in Howell et al. 2001). The main effect of incorporating observational selection effects is to increase the relative contribution of DNe with mass ratios lower than $\sim 0.3$ to the DN CV population.

\section{CB disk detectability}

As in Paper~I, the population synthesis calculations also allow us to
predict the typical properties of possible CB disks in DN and NL CVs,
such as the fraction $\dot{J}_{\rm CB}/\dot{J}_{\rm tot}$ of the total
orbital angular momentum loss caused by the CB disk and the
mass $M_{\rm CB}$ contained in the CB disk. The distributions of these
quantities as functions of the orbital period are shown in
Figs.~\ref{cbjdot} and~\ref{cbmass}, for population synthesis model~A
and the initial mass ratio distribution $n(q)=1$. Accounting for
observational selection effects tends to increase the relative
contribution of systems with higher CB disk orbital angular momentum
loss rates and CB disk masses, but otherwise does not significantly
alter the distributions displayed in Figs.~\ref{cbjdot}
and~\ref{cbmass}.

The orbital angular momentum losses caused by the CB disk
dominate the evolution of the systems close to and beyond the period
bounce for both DNe and NLs, with systems near the upper edge of the
period gap typically losing more than 60\% of their orbital angular
momentum to the CB disk. The mass contained in the CB disk tends to
saturate at $M_{\rm CB} \approx 2$--$3 \times 10^{-4}\,M_\odot$, about
two orders of magnitude larger than the typical CB disk masses found
in Paper~I for systems below the period gap. The total angular
momentum $(GMa)^{1/2}$ is higher for binaries above the period gap,
hence much higher mass transfers into the CB disk are needed to
influence the evolution. With the adopted $\delta=3\times 10^{-4}$,
$M_{\rm CB}$ saturates when about a $1 M_\odot$ of material has been
transferred from the donor star.

Nova-like CVs are probably the best candidates to search for CB disks as most
systems are post-bounce with large disks. This is not surprising as
high mass-transfer rates are needed for the accretion disk to be
stable, and high mass-transfer rates are associated with large angular
momentum losses. Therefore, typical CB disk masses tend to be somewhat
heavier for NLs ($M_{\rm CB} \approx 6$--$20 \times 10^{-5}\,M_\odot$)
than for DNe ($M_{\rm CB} \approx 2$--$8 \times
10^{-5}\,M_\odot$). The CB disks in NLs moreover have a minimum mass
of a few times $10^{-5}\,M_\odot$.

Circumbinary disks around NLs in the 3-5~hr period range should have optically
thick sizes of a few AU, inner surface densities of a few $10^3$
g~cm$^{-2}$, and inner effective temperatures of 1000-2000~K (Taam et
al. 2003). Searches for IR excess (up to $10 \mu$m) or UV
absorption have found no evidence for circumbinary material in the NLs
IX Vel, V592 Cas, QU Car and RW Tri (Belle et al. 2004; Dubus et
al. 2004). However, CB disks are cold and geometrically thin so that
UV absorption requires very favourable line-of-sights (high
inclinations). Furthermore, the cold CB disk dominates the binary
emission only at wavelengths longer than 10\,$\mu$m, so that mid-IR
observations may not be sufficient to uncover them.

\section{Discussion and Conclusions}

The population synthesis of non-magnetic CVs with periods less than 2.75\,hr 
carried out in Paper~I has been extended to non-magnetic CVs with periods longer than 2.75\,hr. Populations of zero-age CVs have been constructed for a range of different input models, and the secular CV evolution has been calculated using an up-to-date stellar and binary evolution code incorporating orbital angular momentum losses due to gravitational radiation, magnetic braking, mass loss through nova explosions, and CB disks.

Adopting a mass input rate into the CB disk equal to $3 \times 10^{-4}$ times the 
mass transfer rate leads to mass transfer rates that can vary by more than an 
order of magnitude at a given orbital period without the introduction of mass 
transfer cycles (cf. King et al. 1995) or considering the effect of nova 
outbursts (Kolb et al. 2001), specifically hibernation (cf. Shara et al. 1986, 
Martin \& Tout 2005).  These rates are sufficiently high ($\sim 10^{-9} M_{\odot} 
\rm{yr^{-1}}$) such that the degree of the departure from thermal equilibrium in 
the donor star causes systems to undergo a bounce at periods of 3--4\,hr, thus, 
providing an alternative explanation for the upper edge of the period gap than 
the standard disrupted magnetic braking model. In this description, systems 
evolving from longer to shorter orbital periods may spend their entire life time 
as DNe or, after an initial DN phase, become NL as the evolution accelerates 
with increasing CB disk mass. On the other hand, systems evolving from shorter 
to longer orbital periods eventually always become DNe due to  the increase in 
the critical mass transfer rate separating transient from persistent CV systems.  
We note that almost all systems containing He WDs belong to this latter group. 
In addition to the difference in sign of the period derivative between the pre- 
and postbounce systems and the much smaller donor mass at a given orbital 
period, the effective temperature of the mass losing donor could differentiate 
the former from the latter.  Specifically,  donors in prebounce systems are 
characterized by higher effective temperatures than the donors in postbounce systems.  

The occurrence of a period bounce also allows binaries above the period gap to 
reach significantly smaller donor mass to accretor mass ratios than in the 
standard disrupted magnetic braking model. In particular, the mass ratios can 
be sufficiently low ($M_d/M_{\rm WD} \la 0.3$) that systems with donors 
that are not nuclearly evolved can exhibit the superhump phenomena at 
orbital periods above the period gap.  In support of this result, Patterson 
et al. (2005) have recently reported that the success rate for searching for 
positive superhump periods in non-magnetic CVs declines from nearly 100\% for 
short period systems ($P_{\rm orb} \la 2.75$\,hr) to about 50\% for systems 
at $3.1 \pm 0.2$\,hr. Upon comparison with our theoretical models, 
we find about 24\% of systems with $P_{\rm orb} = 3.1 \pm 0.2$\,hr have 
$q < 0.3$. When observational selection effects are taken into account, this 
fraction increases to 49--62\%, depending on the adopted observational 
selection factors (see \S\,\ref{selection}). These bias corrected fractions 
are in good agreement with the fractions derived by Patterson et al. (2005). 
Additional observational support for the occurrence of 
superhumps at longer orbital periods in the context of the CB disk model would 
be provided by the determination of the spectral types of these donors which
should differ from nuclearly evolved stars that have lost mass.
On the other hand, our model calculations also reveal that the tail 
of the distribution of systems characterized by low mass ratios can extend to 
6--7 hrs, whereas so far no positive superhumps were detected in non-magnetic 
systems above 4\,hr. In particular, our models for the intrinsic CV 
population predict 15\% of the systems with $P_{\rm orb} > 4$\,hr to have 
$q < 0.3$. However, taking into account observational selection effects 
increases this fraction to 40--53\%. It is possible that the nondetection of 
these systems reflects the low amplitudes of the luminosity modulations 
resulting from the reduced tidal dissipation in the outer accretion disk of 
long period systems with very low mass donors. 

Patterson et al. (2005) and Knigge (2006) furthermore presented mass and radius 
estimates of observed CV donor stars in eclipsing systems and systems exhibiting 
superhumps. In Fig.~\ref{rdmd}, we compare the masses and radii of CV donor stars 
in the theoretical intrinsic population of DN and NL systems with those of the 
long-period ($P_{\rm orb} > 2.75$\,hr) systems considered by Patterson et al. 
(2005) and Knigge (2006). For the theoretical model, we adopted population 
synthesis model~A and the initial mass ratio distribution $n(q)=1$. We furthermore 
do not distinguish between systems exhibiting superhumps and systems not exhibiting 
superhumps in the theoretical population. Considering the uncertainties in the 
observational determination of donor masses and radii and the omission of 
observational selection effects in the theoretical model, the theoretical 
distribution for $M_d \la 0.5\,M_\odot$ is in satisfactory agreement with 
observation. For $M_d \ge 0.5\,M_\odot$, the agreement is less satisfactory, 
but these systems likely have evolved donor stars (Knigge 2006) which are not 
included in our present population synthesis model. 

The birth rates of zero-age CVs born at periods longer than 2.75\,hr are typically of the order of $10^{-3}\,{\rm yr^{-1}}$, which, depending on the adopted initial mass ratio distribution and population synthesis model parameters, is up to an order of magnitude larger than the birth rates found in Paper~I for CVs born at periods less than 2.75\,hr. If the initial secondary mass is assumed to be distributed independently according to the same initial mass function as the primary mass, the number of CVs formed above 2.75\,hr is up to a factor of two smaller than the number of CVs formed below 2.75\,hr. Despite the generally larger birth rates of systems above 2.75\,hr, their shorter life time due to the higher mass transfer rates results in a smaller number of systems above than below the gap in the theoretical present-day CV populations. In particular, our models predict the present-day population of Galactic CVs to consist of $10^5$--$10^6$ systems with periods longer than 2.75\,hr, and $10^6$--$10^7$ systems with period less than 2.75\,hr (Paper~I). The number of systems in the two populations thus differs by about an order of magnitude, which is considerably more than the difference in the number of non-magnetic CVs below and above 2.75\,hr in the observed population: the 2006 January edition 
(RKCat 7.6) of the Ritter \& Kolb (2003) catalog contains $\sim 200$ non-magnetic CVs below 2.75\,hr and $\sim 170$ above 2.75\,hr. As systems below and in the period gap have significantly lower mass transfer rates than systems above the period gap, this possibly reflects stronger observational selection effects operating on systems with periods less than 2.75\,hr. 

The calculations presented in the previous sections are all based on a 
constant magnetic braking time scale of 10\,Gyr. Shorter magnetic braking 
time scales will decrease the time required for post-CE binaries to evolve 
into a CV, as well as increase the longest possible post-CE orbital period 
for which binaries are able to evolve into a CV within the life time of the 
Galaxy. Both of these effects increase the formation rates of ZACVs as a function 
of orbital period. The secular evolution during the CV phase is  
not only affected during the early stages of mass transfer, but also during 
the phase when the CB disk dominates the evolution as 
it has been built up more rapidly. In order to assess the effects of a 
stronger magnetic braking rate on the theoretical DN and NL orbital period 
distributions without recalculating all $\sim 1300$ evolutionary tracks used in 
the population synthesis calculations, we rescaled the orbital period evolution 
of the existing tracks to a stronger magnetic braking rate with a characteristic time scale of $4 \times 10^9$\,years by accelerating the time evolution of the tracks by a factor of $1 + 0.5\, ( 1 - 0.5\,\dot{J}_{\rm CB}/\dot{J}_{\rm tot})$. 
The acceleration factor was determined empirically by comparing a limited number of evolutionary tracks based on magnetic braking time scales of $4 \times 10^9$ years with the corresponding tracks based on magnetic braking time scales of $10^{10}$ years. As illustrated in Fig.~\ref{accel}, application of the multiplication factor leads to evolutionary tracks that are very close to the evolutionary tracks calculated for a magnetic braking time scale of $4 \times 10^9$ years.  The resulting orbital period distributions of the intrinsic DN and NL populations are shown in the left- and right-hand panels of Fig.~\ref{pintMB2}, respectively. The stronger magnetic braking rate accelerates the evolution and increases the mass transfer rates during the early stages of CV evolution. 
As a consequence, the relative number of DNe near the upper edge of the period gap is reduced, while the relative number of NLs near the upper edge of the period gap is increased (cf. Fig.~\ref{pint}). The overall agreement of the theoretical with the observed distributions therefore improves, suggesting that magnetic braking time scales are shorter than $10^{10}$\,years. 

\begin{figure}
\center
\resizebox{8.0cm}{!}{\includegraphics{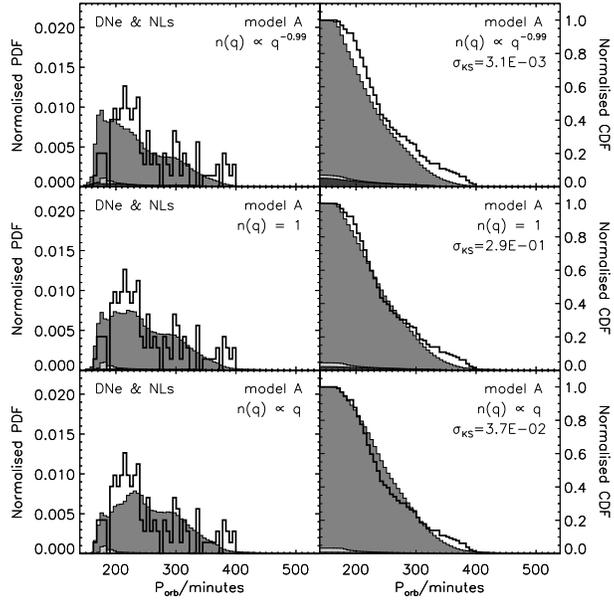}}
\caption{Orbital period distribution of non-magnetic CVs with periods
  longer than 2.75\,hr for population synthesis model A and different
  initial mass ratio distributions $n(q)$, without regard for
  observational bias. The light, intermediate, and dark gray shading
  represents the fractions of He, C/O, and O/Ne/Mg WD systems
  contributing to the population (note that the dark gray is only clearly visible in the top right panel). The thick solid line represents the
  observed CV orbital period distribution. For ease of comparison, the
  PDFs and CDFs are normalized to unity. The degree of agreement
  between the observed and simulated CDFs is indicated by the
  Kolmogorov-Smirnov significance level $\sigma_{\rm KS}$.}
\label{pint0}
\end{figure}

\begin{figure*}
\resizebox{8.0cm}{!}{\includegraphics{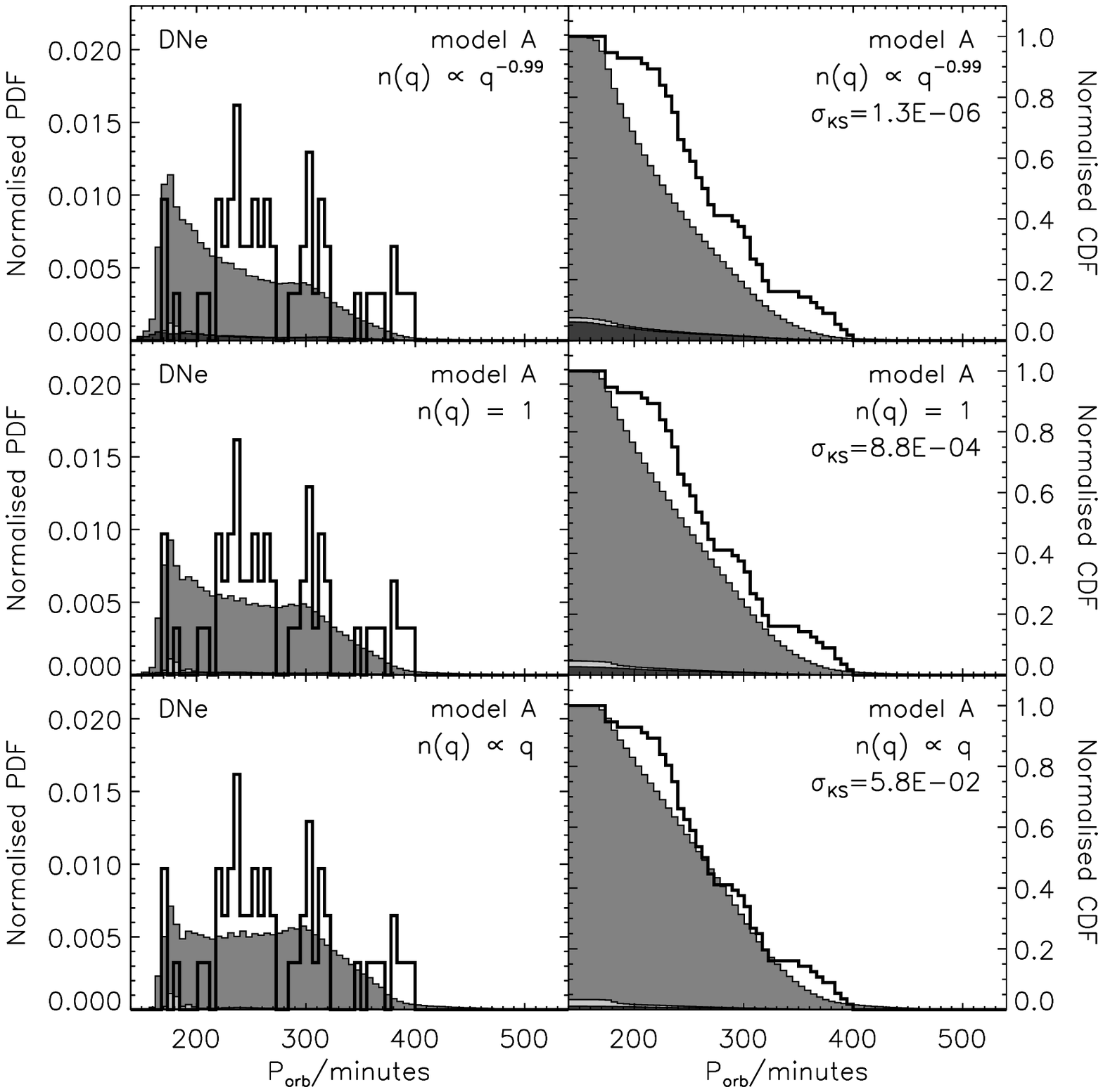}} \hspace{0.2cm}
\resizebox{8.0cm}{!}{\includegraphics{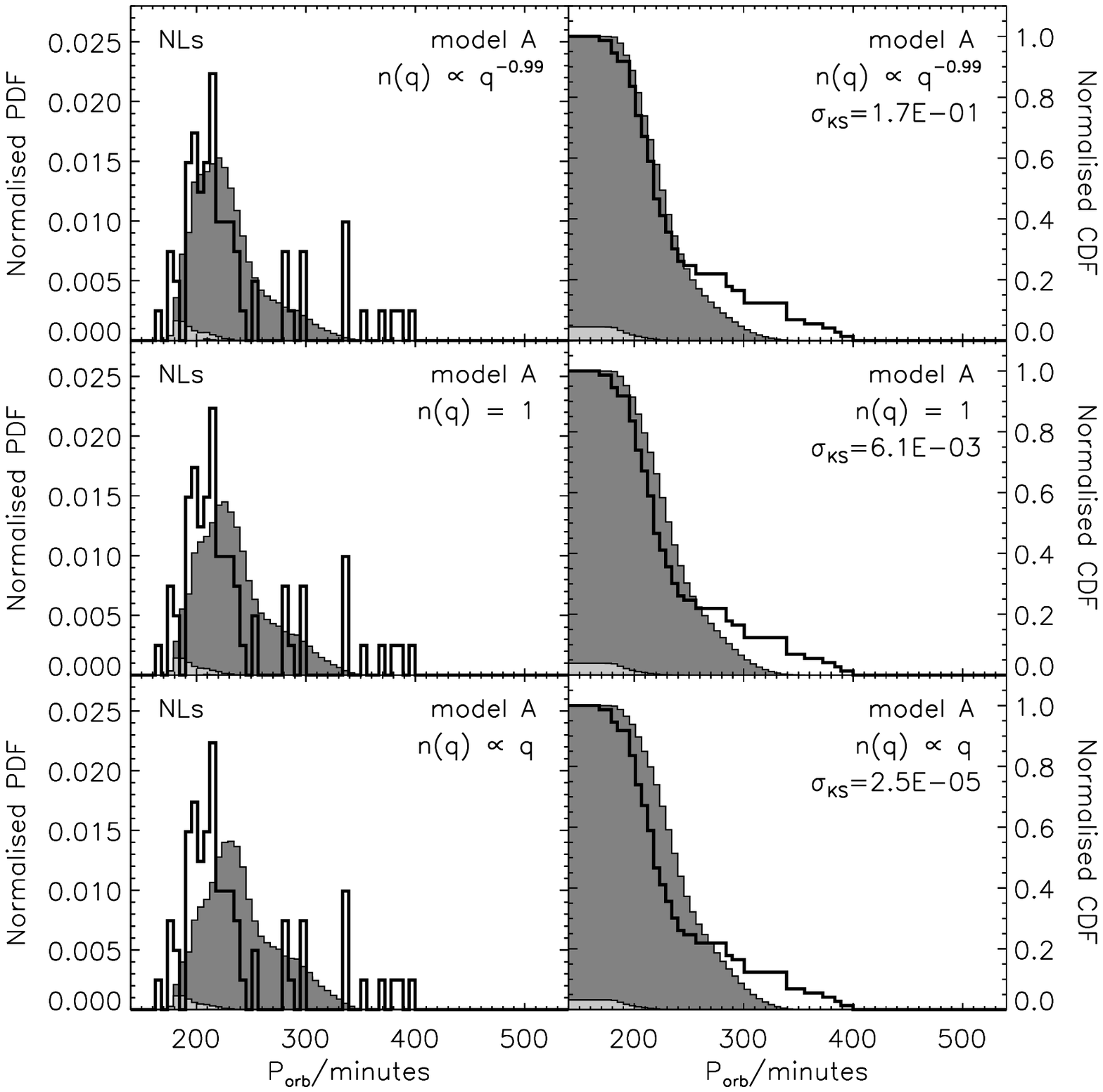}} \hspace{0.2cm}
\caption{Same as Fig.~\ref{pint0}, but with the population of CVs divided into DNe (left) and NLs
  (right).}
\label{pint}
\end{figure*}

\begin{figure}
\center
\resizebox{8.0cm}{!}{\includegraphics{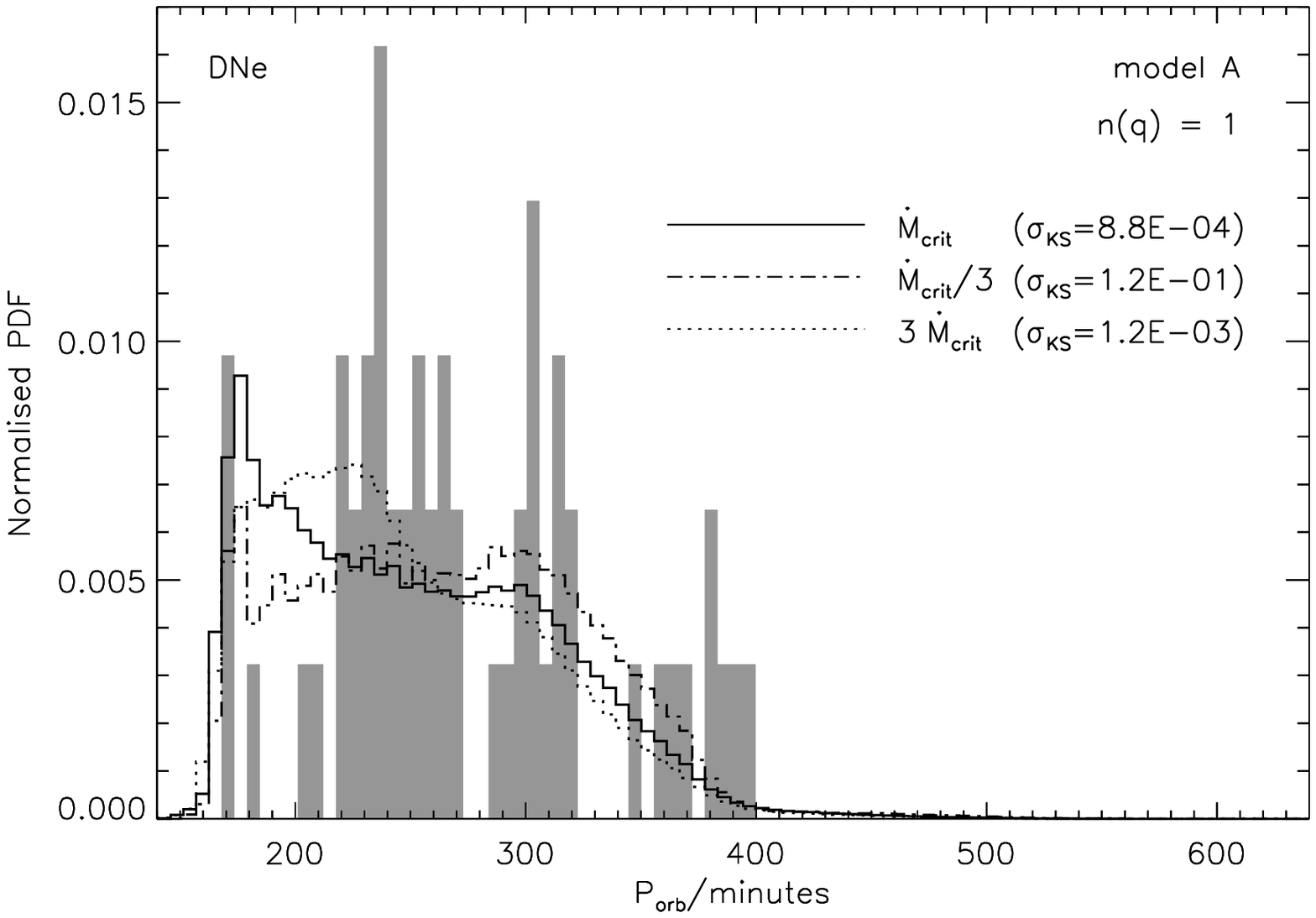}} \hspace{0.2cm} \\
\resizebox{8.0cm}{!}{\includegraphics{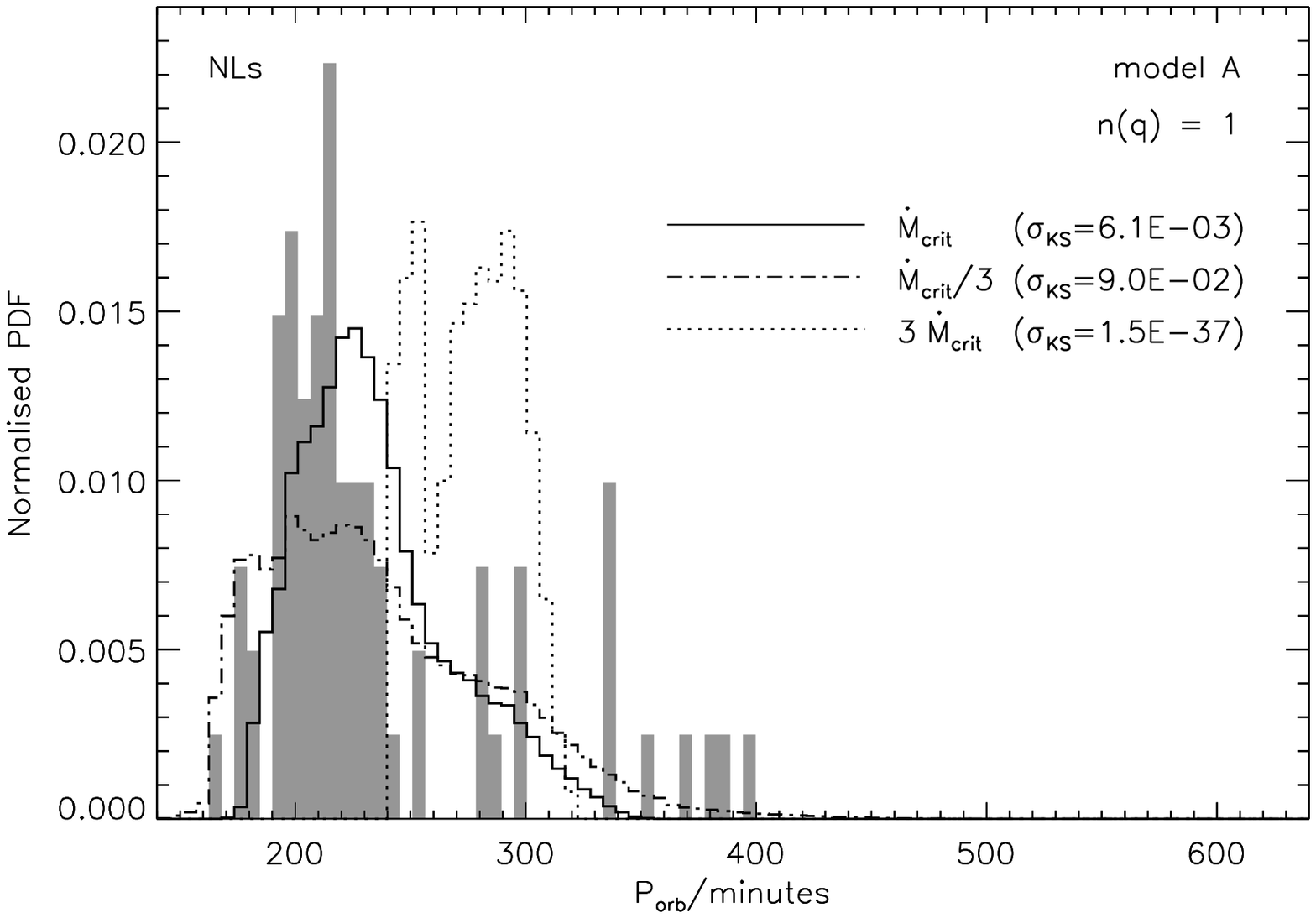}} \hspace{0.2cm}
\caption{Normalized intrinsic orbital period distributions of DNe (top) and NLs 
(bottom) with periods above 2.75\,hr, for population synthesis model A, the 
initial mass ratio distribution $n(q)=1$, and different critical mass transfer 
rates $\dot{M}_{\rm crit}$ separating transient from persistent systems in 
the CV population synthesis. The solid lines represent the orbital period 
distributions for critical mass transfer rates given by Eq.~(\ref{mdotcrit}), 
while the dash-dotted and dotted lines respectively represent the orbital 
period distributions for critical mass transfer rates that are a factor of 3 
smaller and a factor 3 larger than those given by Eq.~(\ref{mdotcrit}). The 
observed DNe and NL orbital period distributions are represented by the grey 
shaded histograms.}
\label{mdc}
\end{figure}

The population synthesis calculations used to construct the zero-age CV populations incorporate a stability test of mass transfer when the donor first fills its critical Roche lobe.  Systems which do not satisfy the test of stability are not followed further and are removed from the population. The stability test is 
performed in accord with the fully non-conservative approximation adopted in the calculation of the binary evolutionary tracks. However, once mass transfer is 
initiated, it takes a finite amount of time to build up the accretion disk and 
evolve towards the first nova explosion. Thus, matter is transferred to the WD  accretor and the conservative mass transfer stability criterion may therefore be more appropriate than the non-conservative one in the construction of the ZACV population. In order to illustrate the effects of the stability criterion, the theoretical orbital distribution of non-magnetic CVs obtained assuming conservative mass transfer for the construction of the zero-age CV population is shown in Fig.~\ref{pintC}. From comparison with Fig.~\ref{pint0}, 
it follows that the main effect of the conservative mass transfer stability criterion 
is to introduce a small bump in the theoretical orbital period distribution which 
coincides with the secondary peak at $P_{\rm orb} \simeq 300$\,minutes in the 
observed orbital period distribution. This is particularly striking for the initial 
mass ratio distribution $n(q) \propto q$. The  conservative mass transfer stability 
criterion also increases the relative contribution of systems with O/Ne/Mg WDs for 
which stability is less of an issue (because the stability of conservative mass 
transfer requires smaller $M_d/M_{\rm WD}$ ratios than non-conservative mass transfer). Similar tendencies are observed when the DN population is considered separately.  The main effect on the NL population is to increase the peak in the theoretical distribution at $P_{\rm orb} \simeq 220$\,minutes and shift the cut-off near the upper edge of the period gap from 180 to 200\,minutes, which is at somewhat too long of an orbital period compared to the cut-off in the observed orbital period distribution.

\begin{figure*}
\center
\resizebox{8.0cm}{!}{\includegraphics{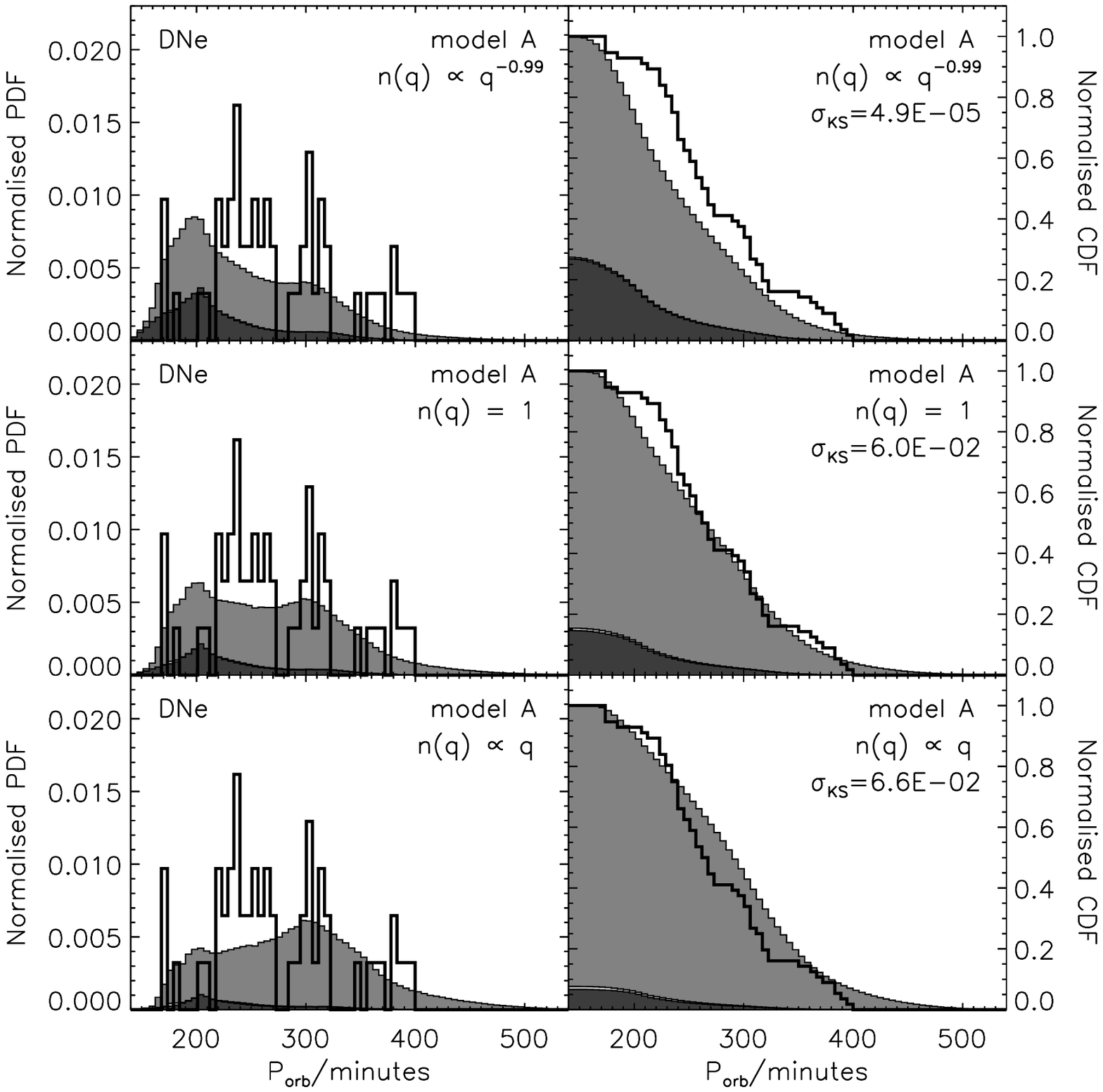}} 
\resizebox{8.0cm}{!}{\includegraphics{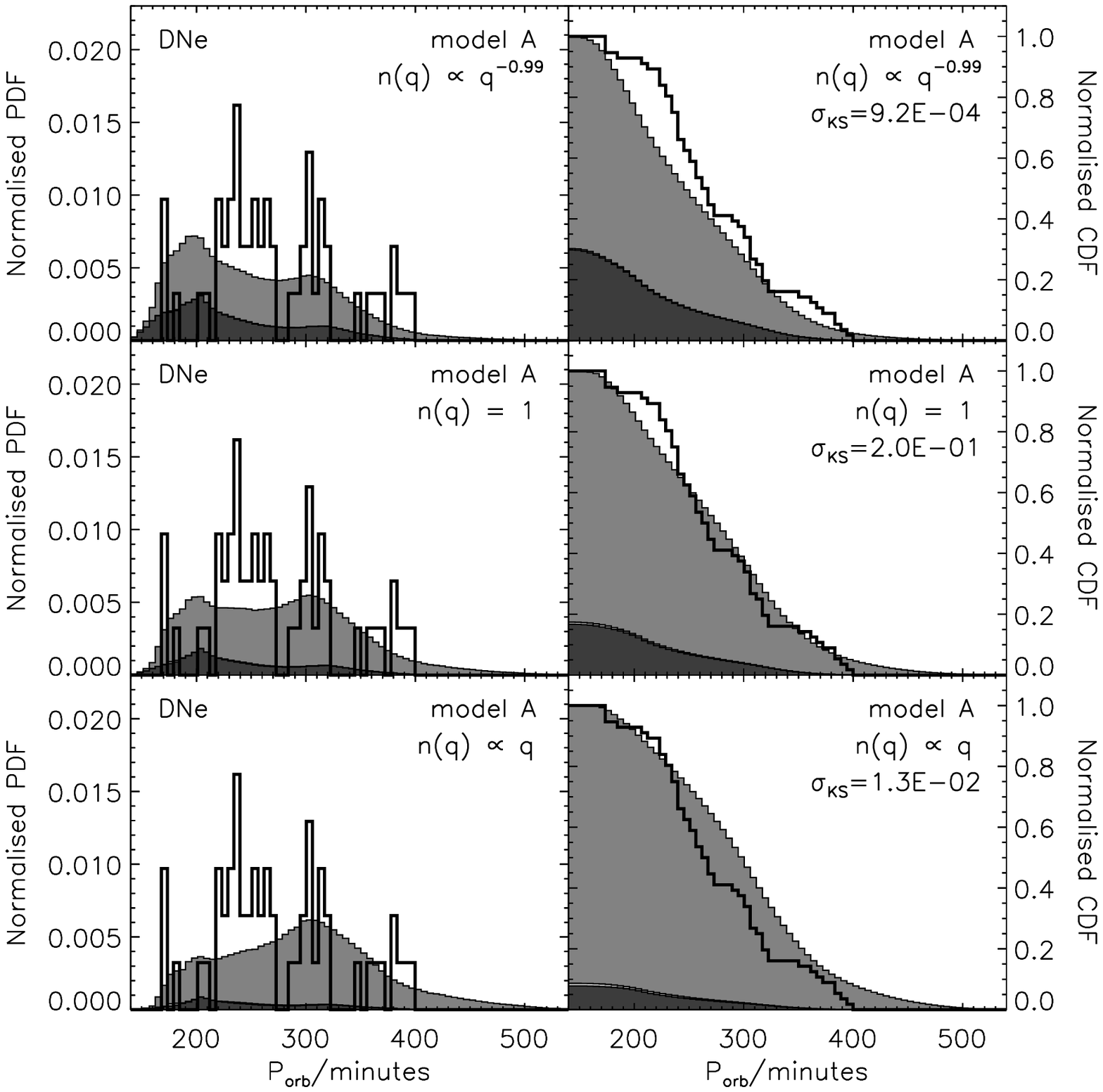}} \\
\resizebox{8.0cm}{!}{\includegraphics{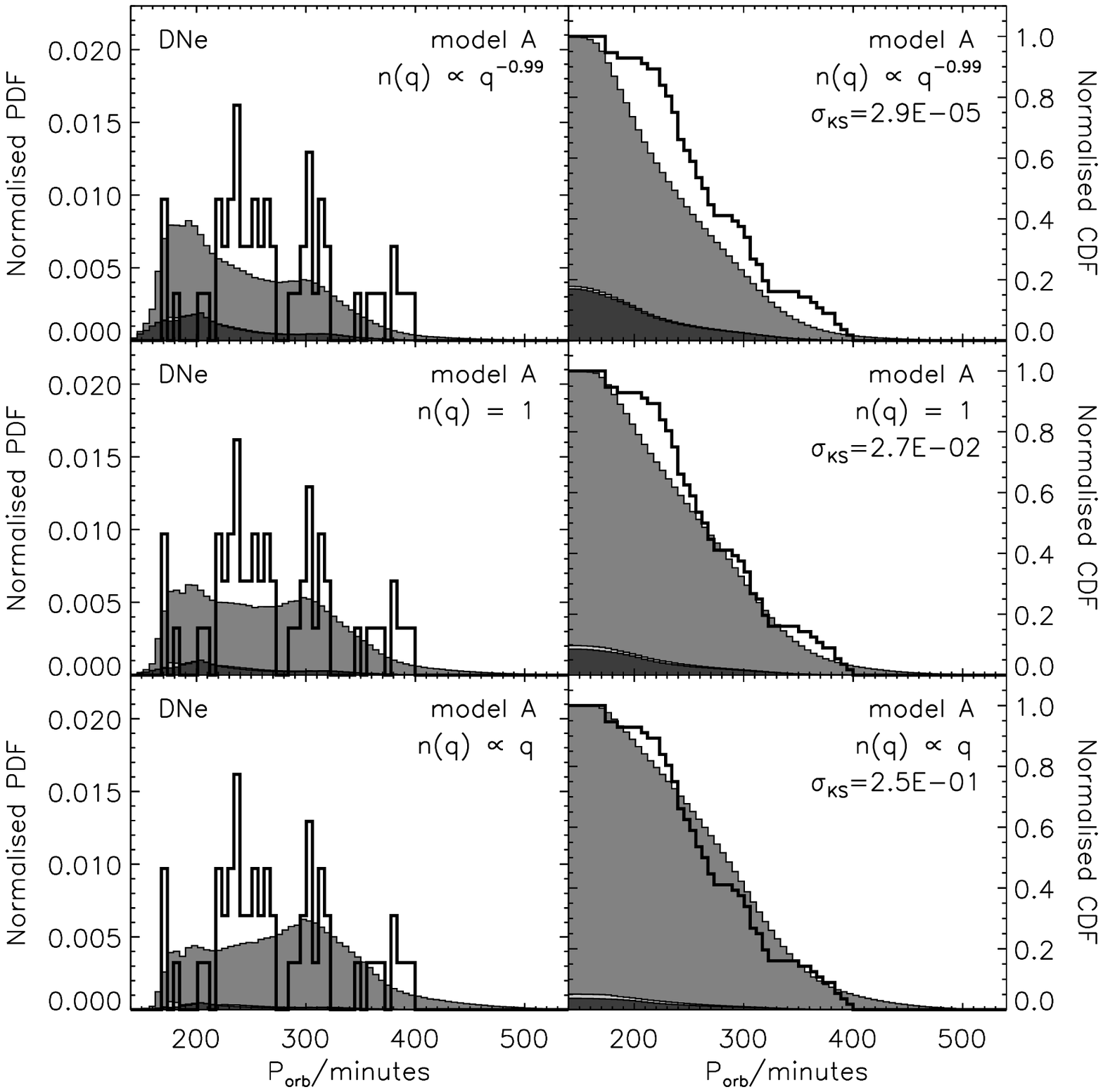}} \vspace{0.2cm} 
\resizebox{8.0cm}{!}{\includegraphics{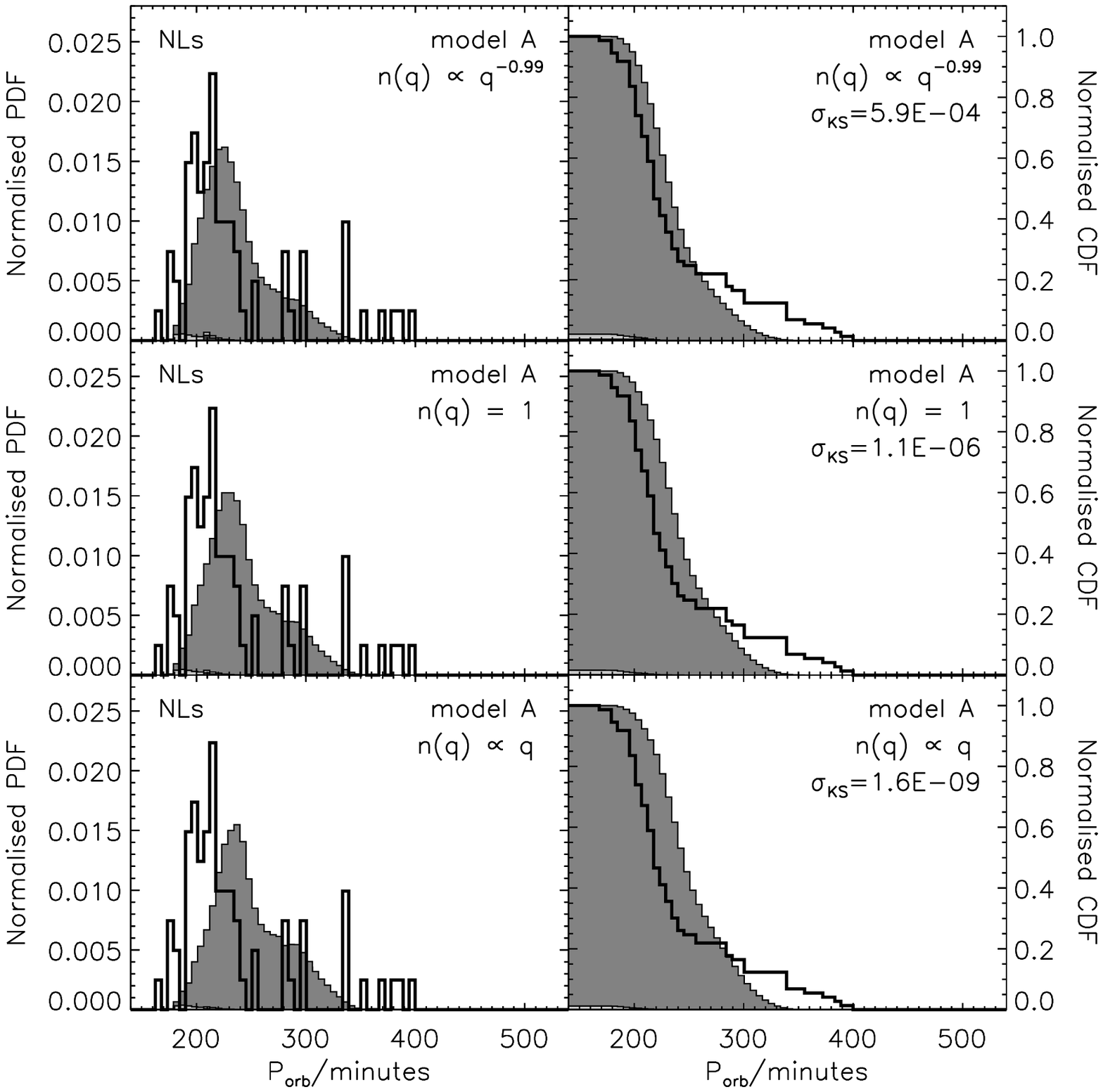}}
\caption{Same as Fig.~\ref{pint}, but corrected for observational bias. Selection effects for DNe are modeled using a detection probability factor $W=d\,L_{\rm acc}^{1.5}$ with $d=0.1$ (top left panel), $W=d\,L_{\rm acc}^{1.5}$ with $d=\dot{M}_d/\dot{M}_{\rm crit}$ (top right panel), or $W=d\,L_{\rm acc}$ with $d=\dot{M}_d/\dot{M}_{\rm crit}$ (bottom left panel). Selection effects for NLs (bottom right panel) are modeled using a detection
probability factor $W=L_{\rm acc}$ with $d=1$. Note that from Eqs.~(\ref{d1}) 
and~(\ref{d2}) it follows that  $W=d\,L_{\rm acc}$ with $d=\dot{M}_d/\dot{M}_{\rm crit}$ 
and $W=L_{\rm acc}$ with $d=1$ both reduce to $W=GM_{\rm WD}\,\dot{M}_d/R_{\rm WD}$.}
\label{pwgt}
\end{figure*}

\begin{figure*}
\resizebox{8.0cm}{!}{\includegraphics{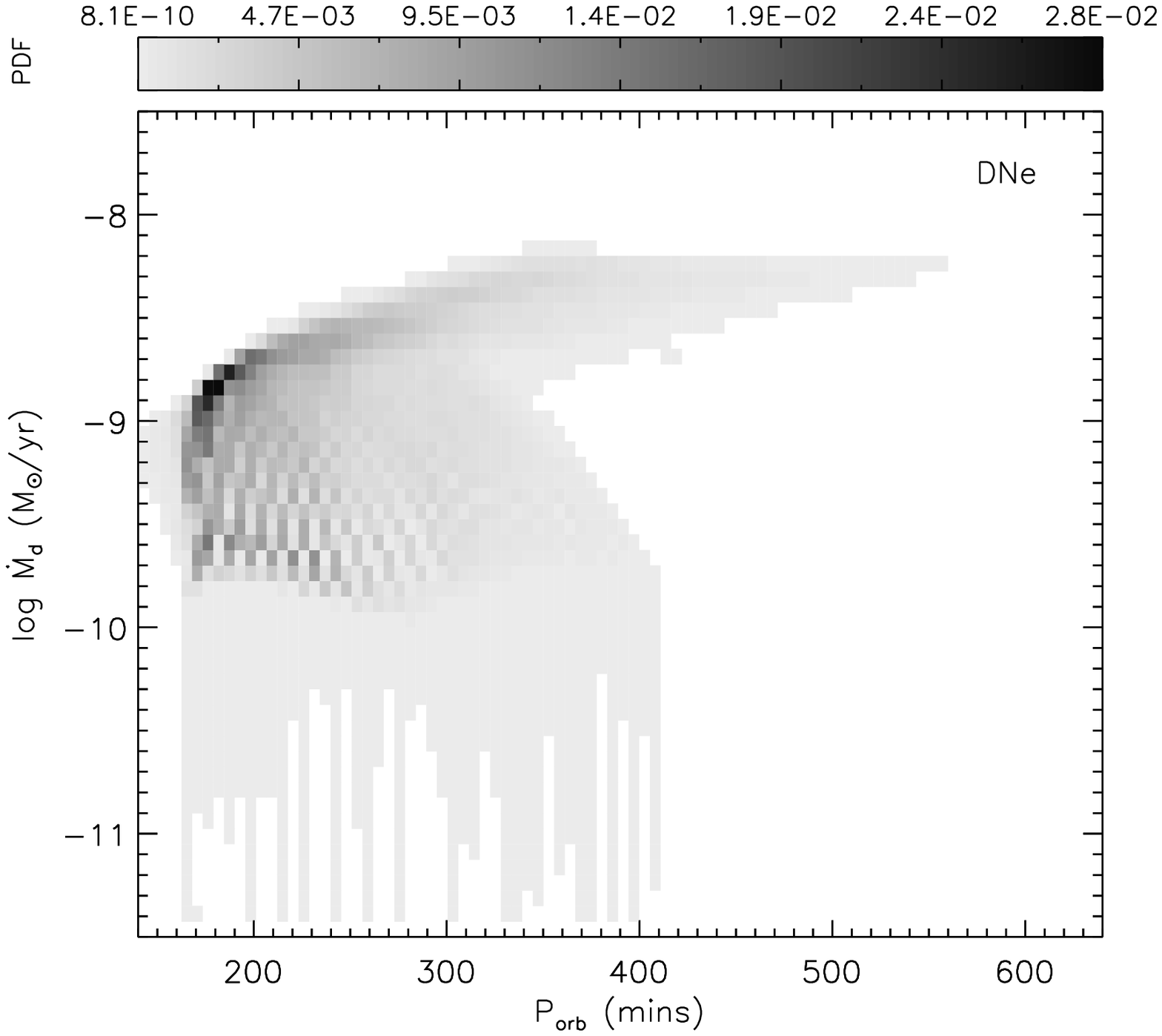}} \hspace{0.2cm}
\resizebox{8.0cm}{!}{\includegraphics{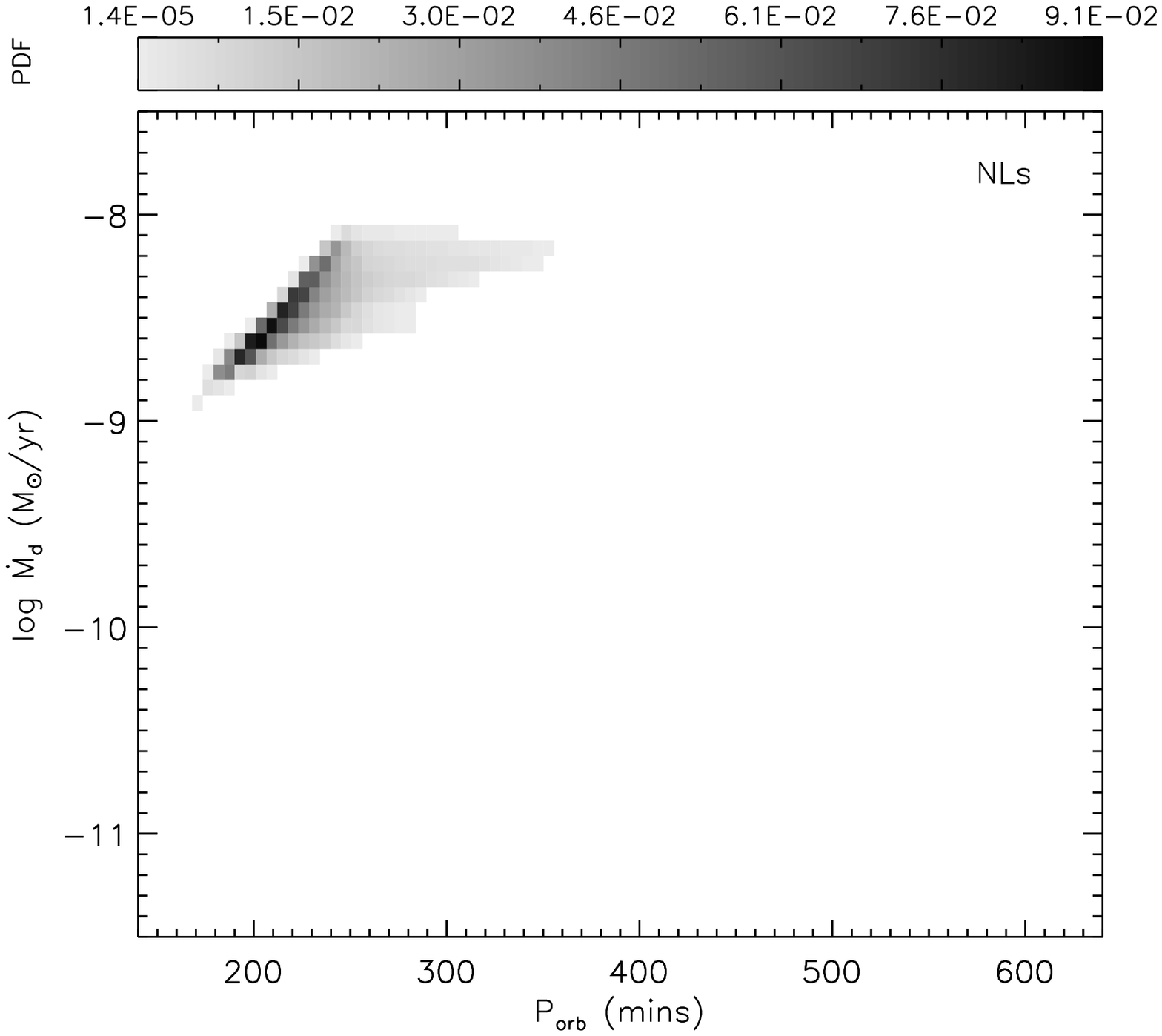}} \hspace{0.2cm}
\caption{Normalized distribution of DN (left) and NL (right)
  mass-transfer rates as a function of the orbital period for
  population synthesis model A and the initial mass ratio distribution
  $n(q)=1$, without regard for observational bias. The ``fan" of high- and low-density lines at mass-transfer rates below $10^{-9}\,M_\odot\,{\rm yr^{-1}}$ in the DN population is due to the finite number of evolutionary tracks underlying the calculation.}
\label{mtint}
\end{figure*}

\begin{figure*}
\resizebox{8.0cm}{!}{\includegraphics{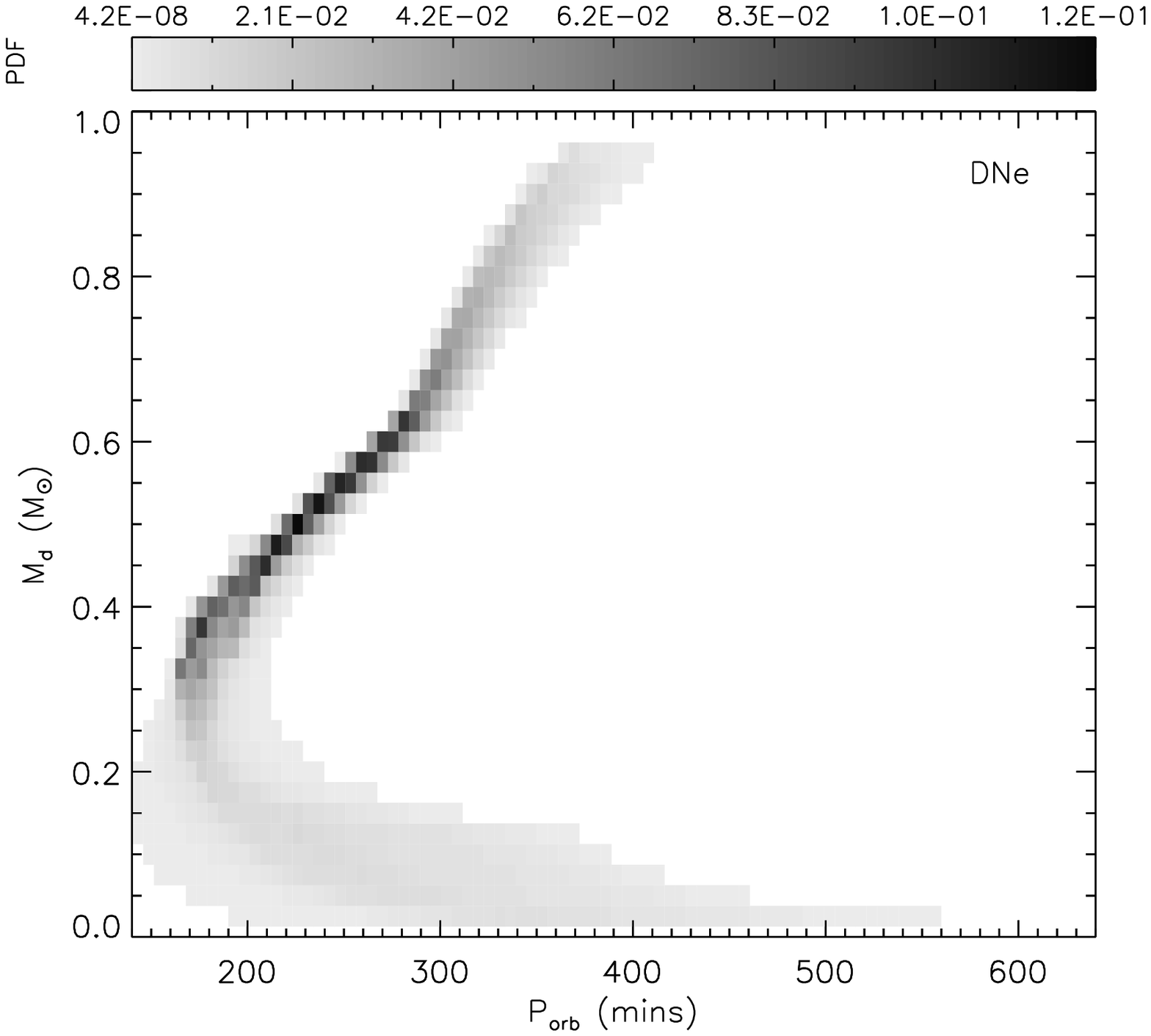}} \hspace{0.2cm}
\resizebox{8.0cm}{!}{\includegraphics{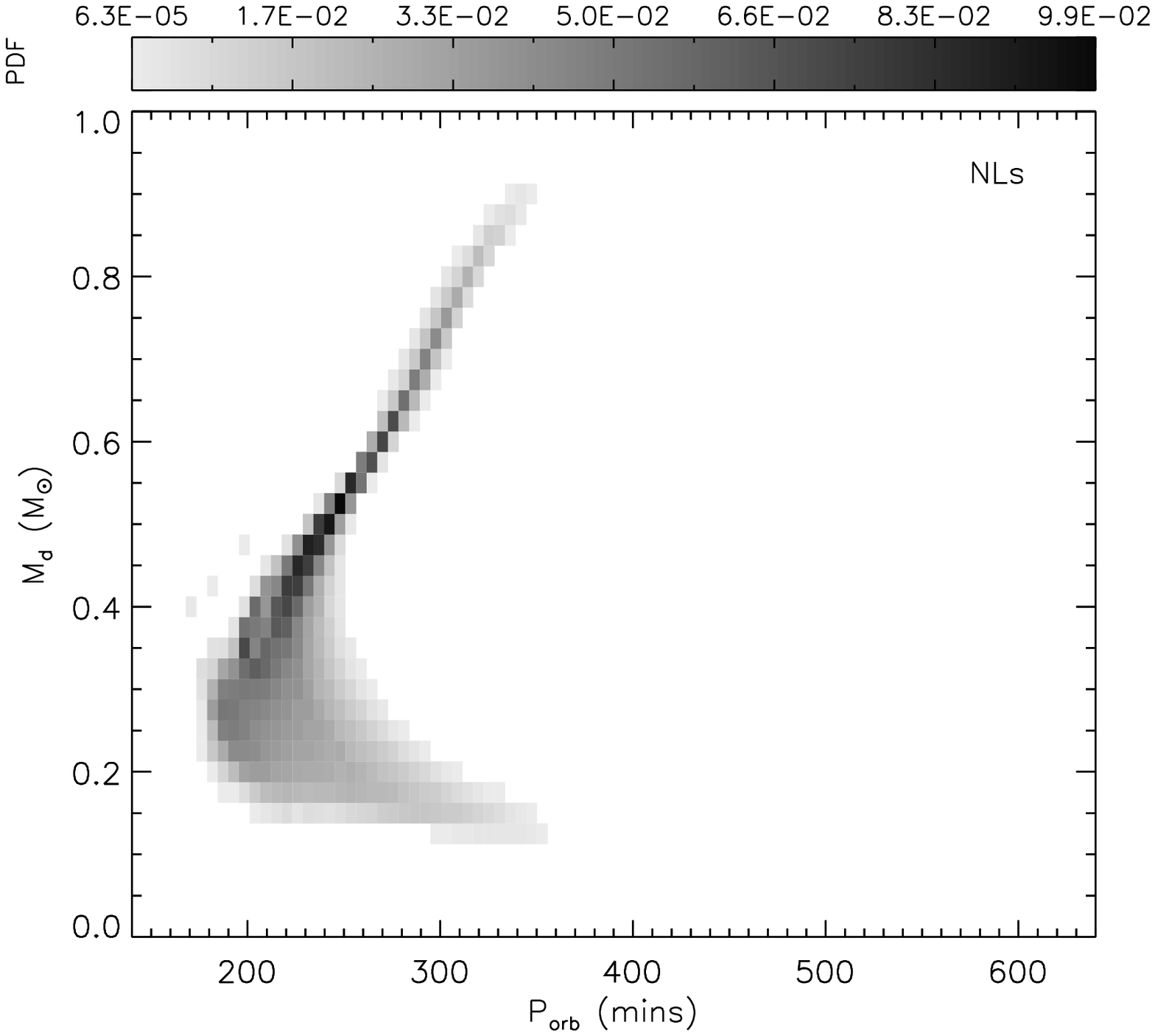}} \hspace{0.2cm}
\caption{Normalized distribution of DN (left) and NL (right) donor
  masses as a function of the orbital period for population synthesis
  model A and the initial mass ratio distribution $n(q)=1$, without
  regard for observational bias.}
\label{mdint}
\end{figure*}

\begin{figure*}
\resizebox{8.0cm}{!}{\includegraphics{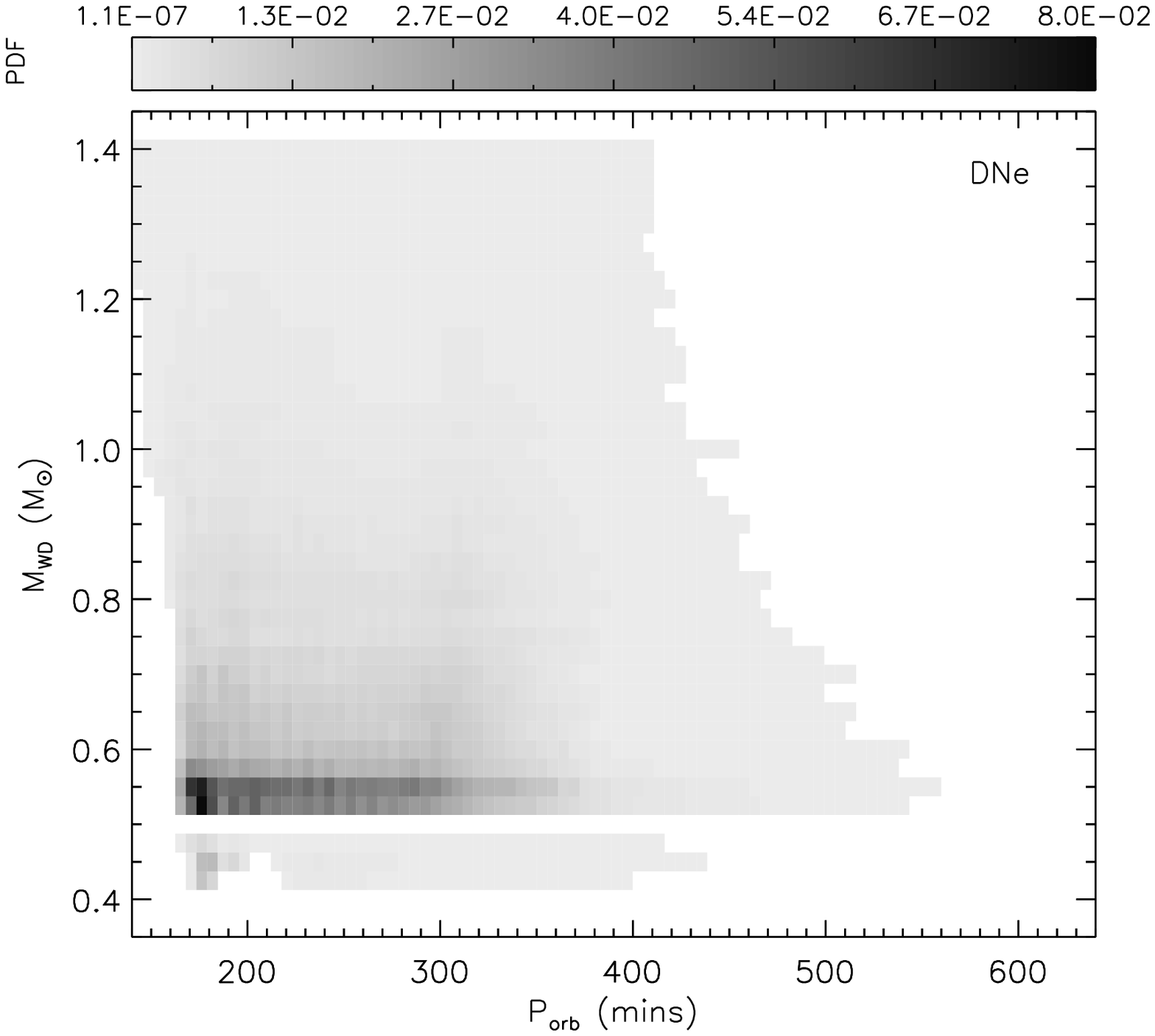}} \hspace{0.2cm}
\resizebox{8.0cm}{!}{\includegraphics{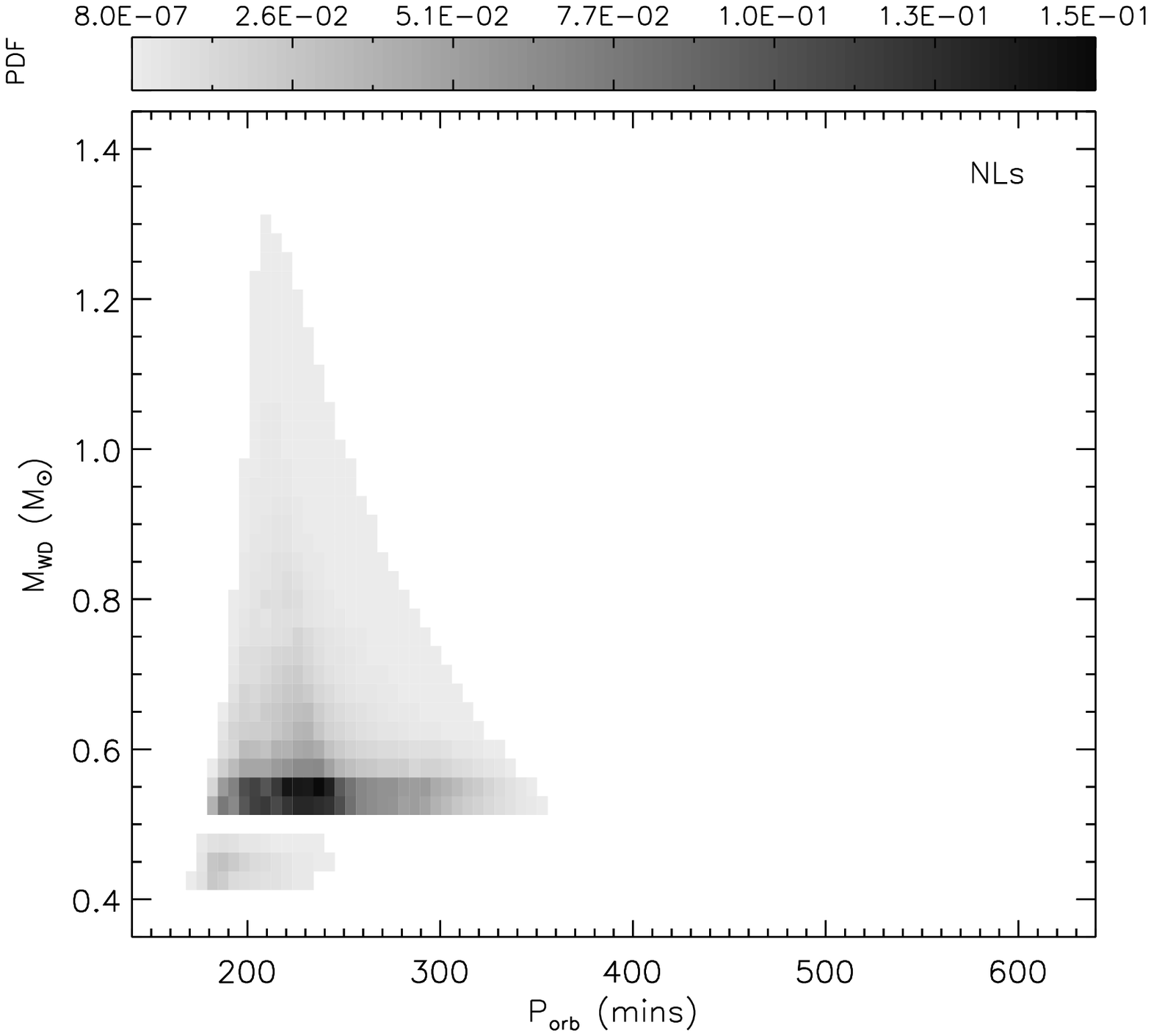}} \hspace{0.2cm}
\caption{Normalized distribution of DN (left) and NL (right) WD
  masses as a function of the orbital period for population synthesis
  model A and the initial mass ratio distribution $n(q)=1$, without
  regard for observational bias.}
\label{mwdint}
\end{figure*}

\begin{figure*}
\resizebox{8.0cm}{!}{\includegraphics{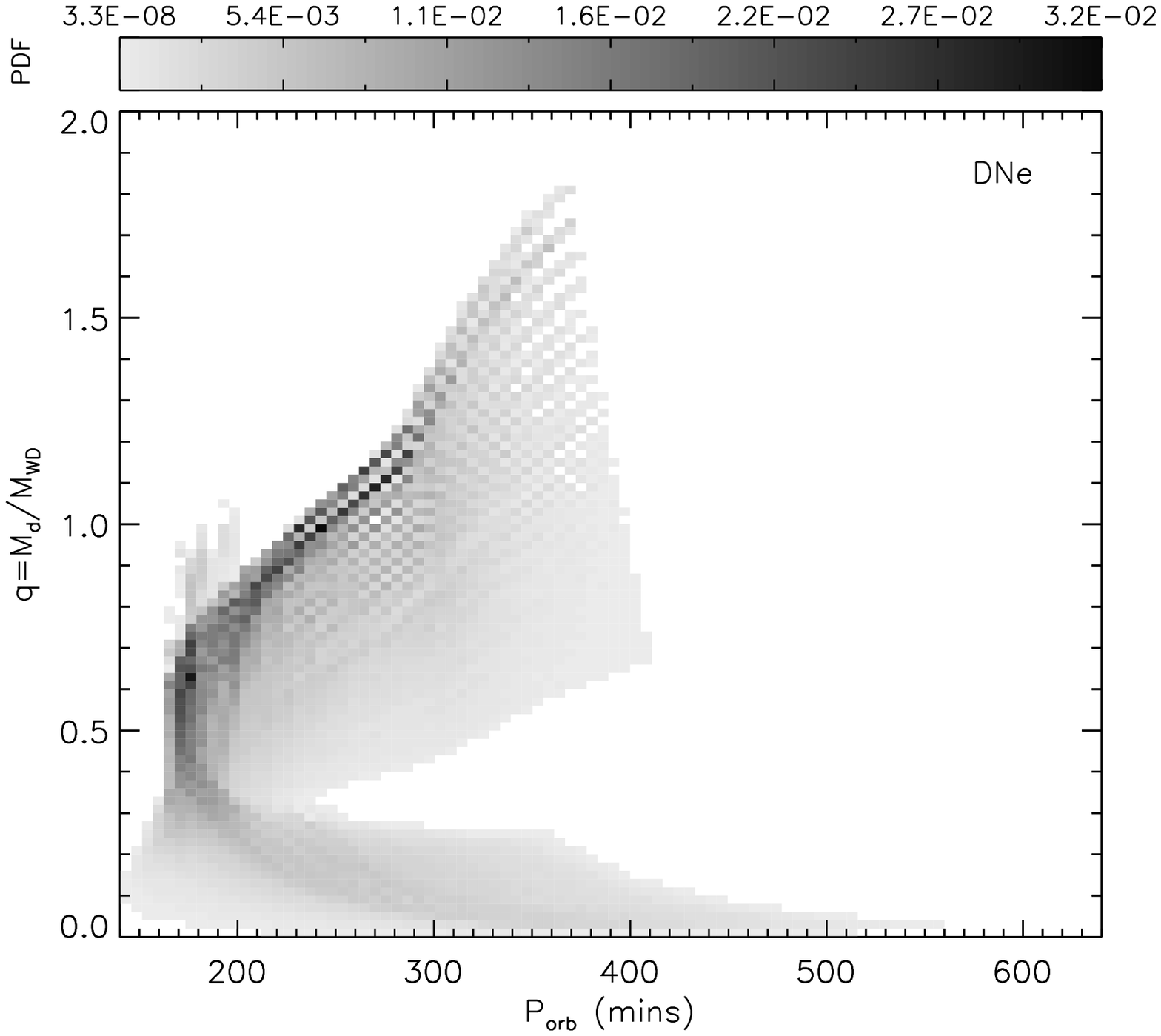}} \hspace{0.2cm}
\resizebox{8.0cm}{!}{\includegraphics{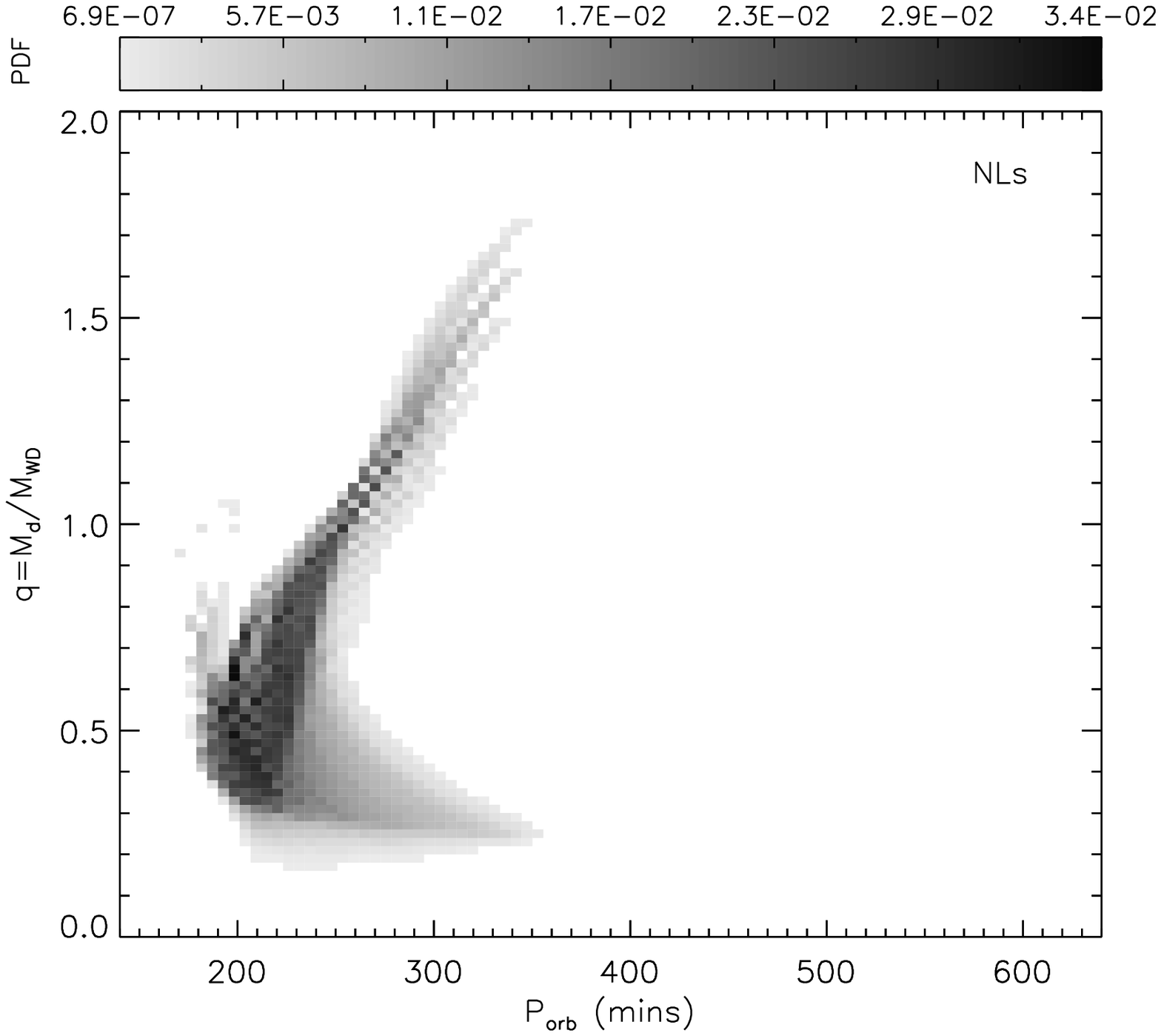}} \hspace{0.2cm}
\caption{Normalized distribution of DN (left) and NL (right) mass
  ratios as a function of the orbital period for population synthesis
  model A and the initial mass ratio distribution $n(q)=1$, without
  regard for observational bias. The ``fan" of high- and low-density lines at mass ratios $q \ga 1$ is due to the finite number of evolutionary tracks underlying the calculation.}
\label{qint}
\end{figure*}

\begin{figure*}
\resizebox{8.0cm}{!}{\includegraphics{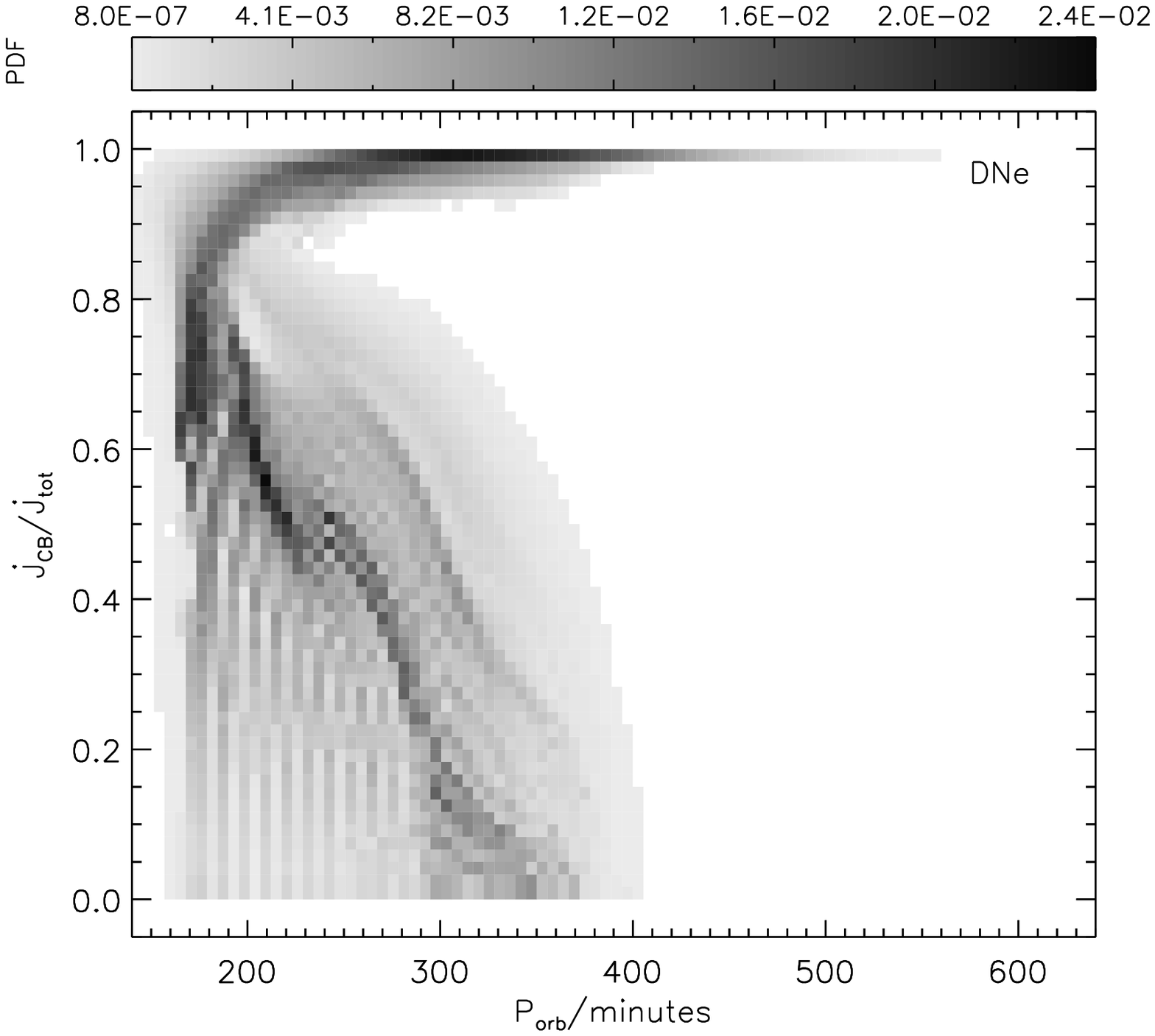}} \hspace{0.2cm}
\resizebox{8.0cm}{!}{\includegraphics{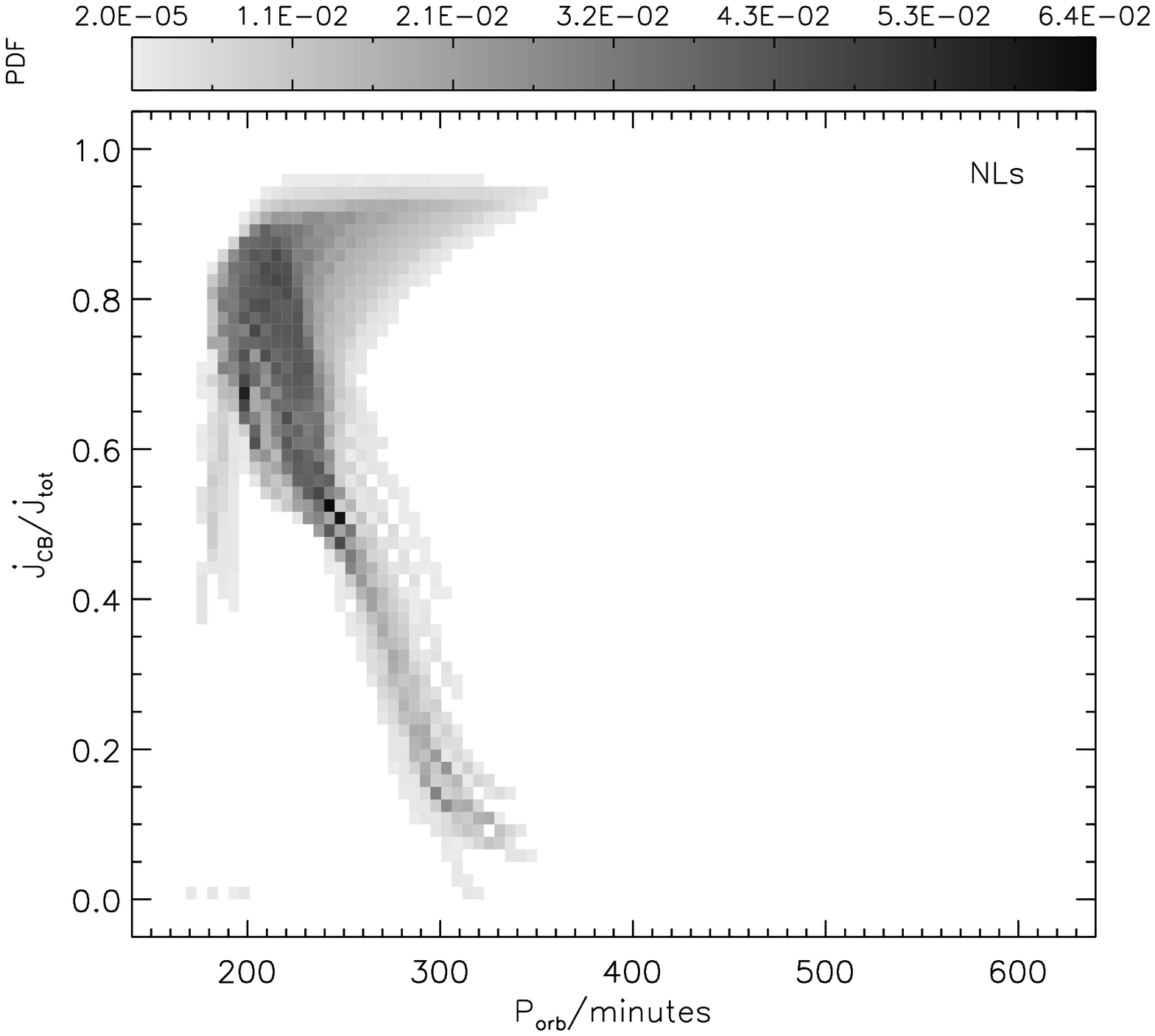}} \hspace{0.2cm}
\caption{Normalized distribution of $\dot{J}_{\rm CB}/\dot{J}_{\rm
    tot}$ for DNe (left) and NLs (right) as a function of the orbital period for
  population synthesis model~A and the initial mass ratio distribution
  $n(q)=1$, without regard for observational bias. The ``fan" of high- and low-density lines at $\dot{J}_{\rm CB}/\dot{J}_{\rm tot} \la 0.6$ is due to the finite number of evolutionary tracks underlying the calculation. } 
\label{cbjdot}
\end{figure*}

\begin{figure*}
\resizebox{8.0cm}{!}{\includegraphics{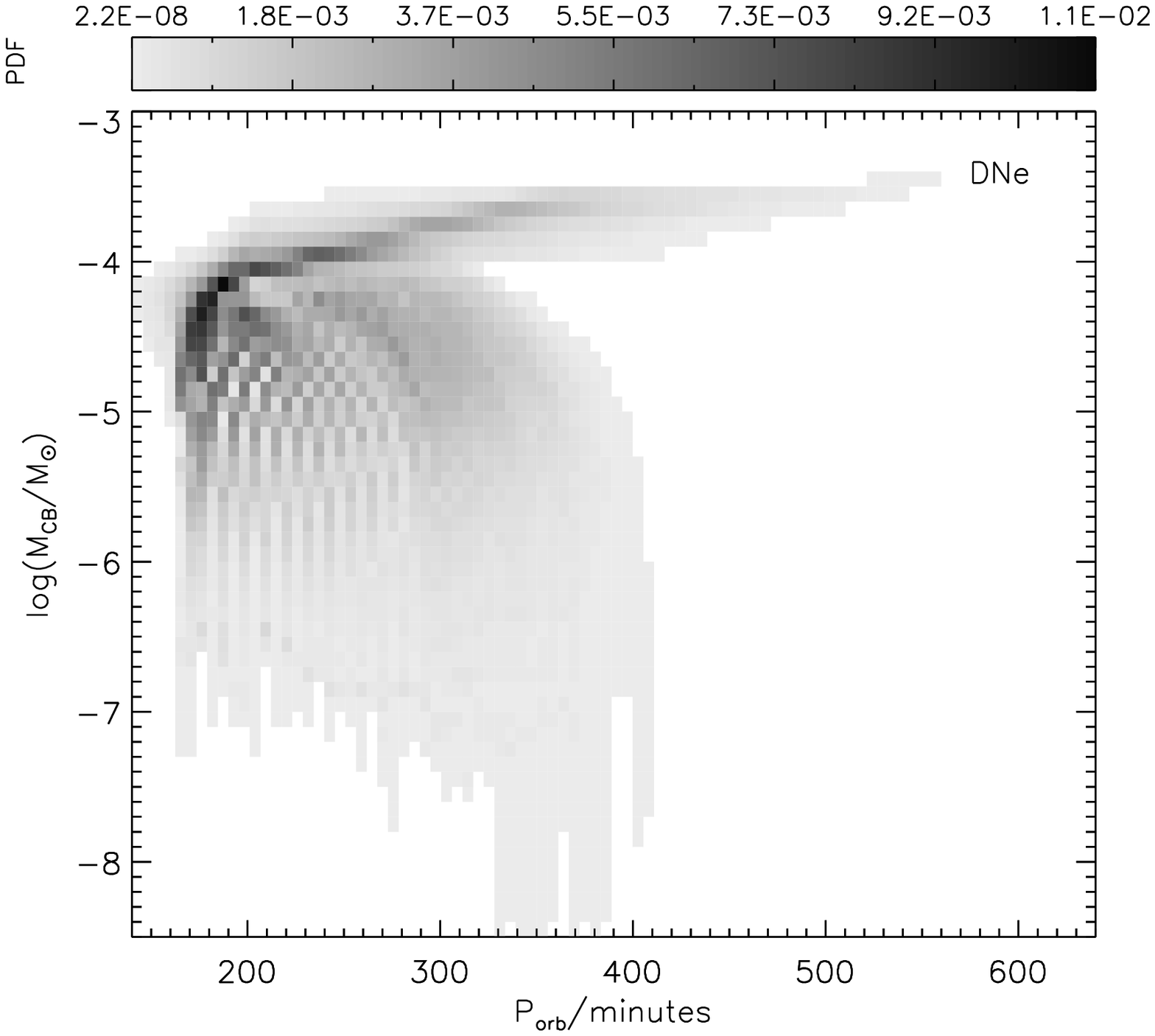}}
\resizebox{8.0cm}{!}{\includegraphics{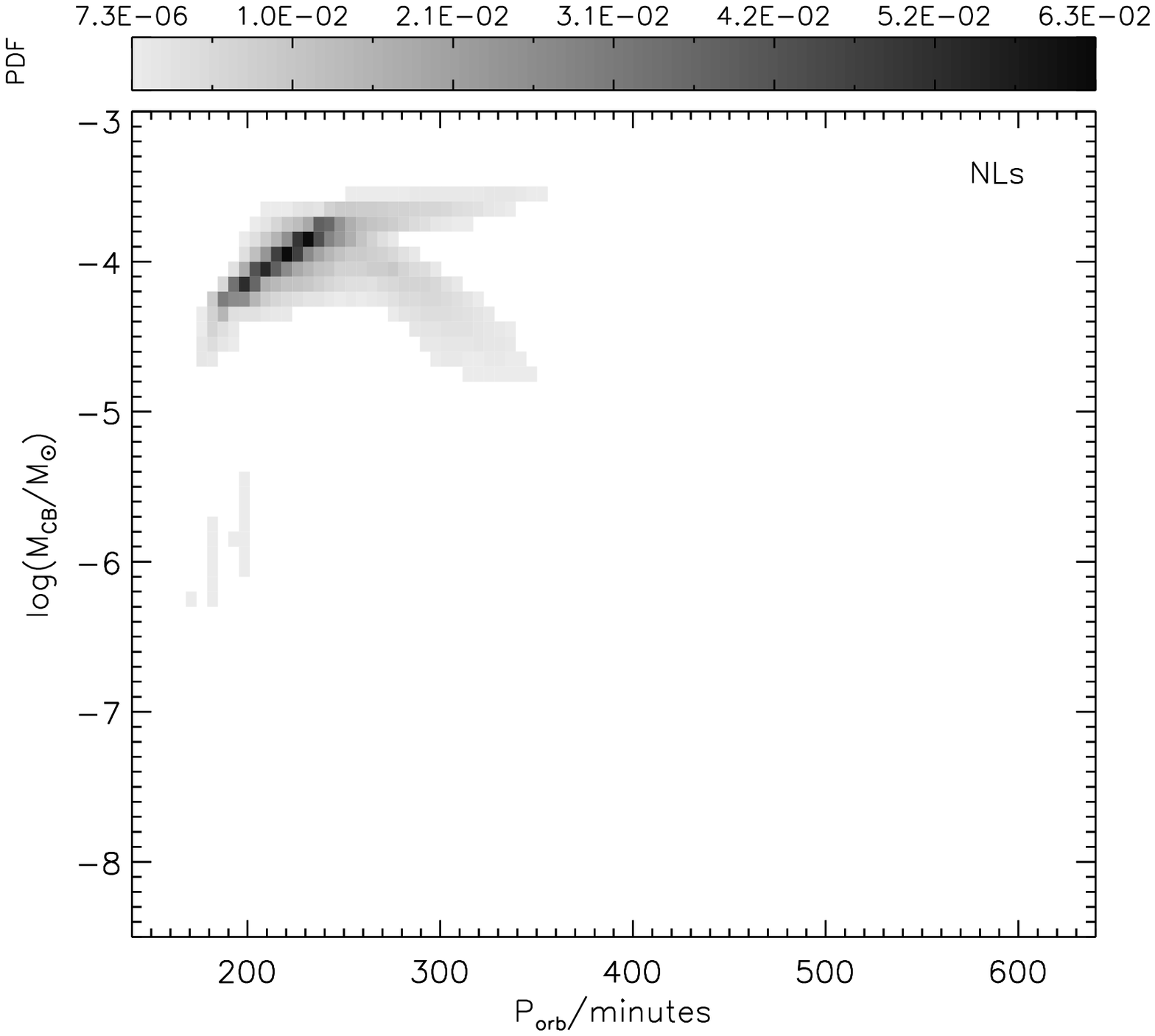}}
\caption{Same as Fig.~\ref{cbjdot}, but for the mass $M_{\rm CB}$ contained in the CB disk. The ``fan" of high- and low-density lines at $\log M_{\rm CB}/M_\odot \la -4.5$ in the DN population is due to the finite number of evolutionary tracks underlying the calculation.}
\label{cbmass}
\end{figure*}

\begin{figure}
\resizebox{8.0cm}{!}{\includegraphics{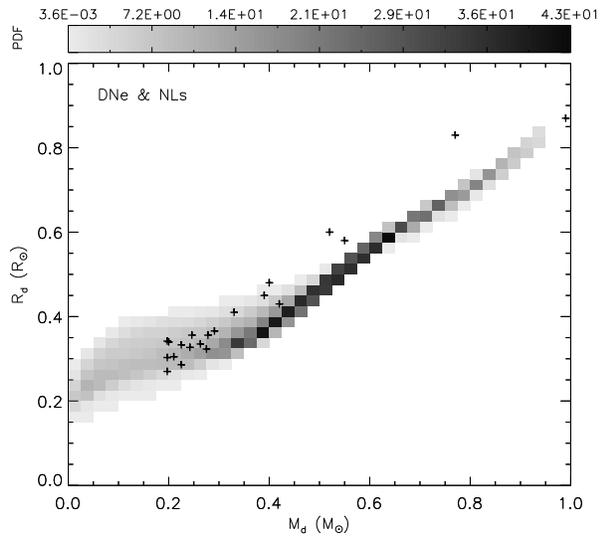}} 
\caption{Normalized distribution of DN and NL CV donor radii 
  as a function of the donor mass for population synthesis
  model A and the initial mass ratio distribution $n(q)=1$, without
  regard for observational bias. Crosses indicate the masses and radii of donor stars in observed CVs with $P_{\rm orb} > 2.75$\,hr [data taken from Table~8 in Patterson et al. (2005) and Table~1 in Knigge (2006)].}
\label{rdmd}
\end{figure}

\begin{figure}
\center
\resizebox{8.0cm}{!}{\includegraphics{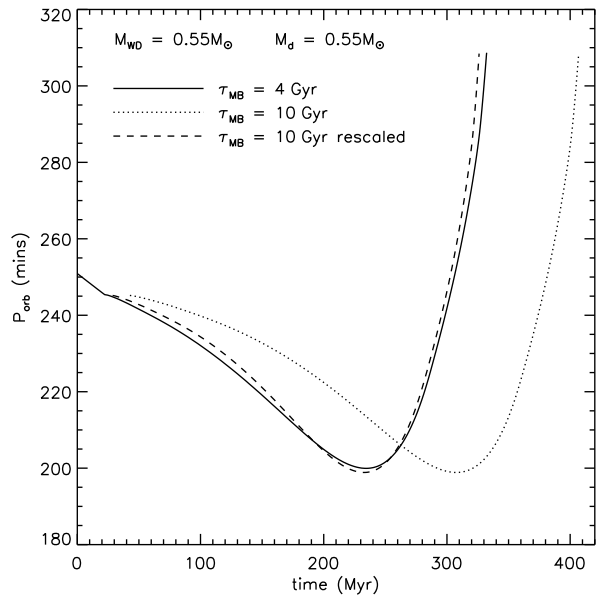}} 
\caption{Time evolution of the orbital period of a CV evolving under the influence of a CB disk and a magnetic braking time scale of $\tau_{\rm MB} = 4 \times 10^9$\,years (solid line) or $\tau_{\rm MB} =10^{10}$\,years (dotted line). The initial WD and donor masses are equal to $0.55\,M_\odot$ in both cases. The dashed line corresponds to the time evolution of the $\tau_{\rm MB} =10^{10}$\,year evolutionary track multiplied by a factor of $1 + 0.5\, ( 1 - 0.5\,\dot{J}_{\rm CB}/\dot{J}_{\rm tot})$. The resulting orbital period evolution is in excellent agreement with the $\tau_{\rm MB} = 4 \times 10^9$\,year evolutionary track, with the minimum attained period differing by less than $\sim 1$\,minute.}
\label{accel}
\end{figure}

\begin{figure*}
\resizebox{8.0cm}{!}{\includegraphics{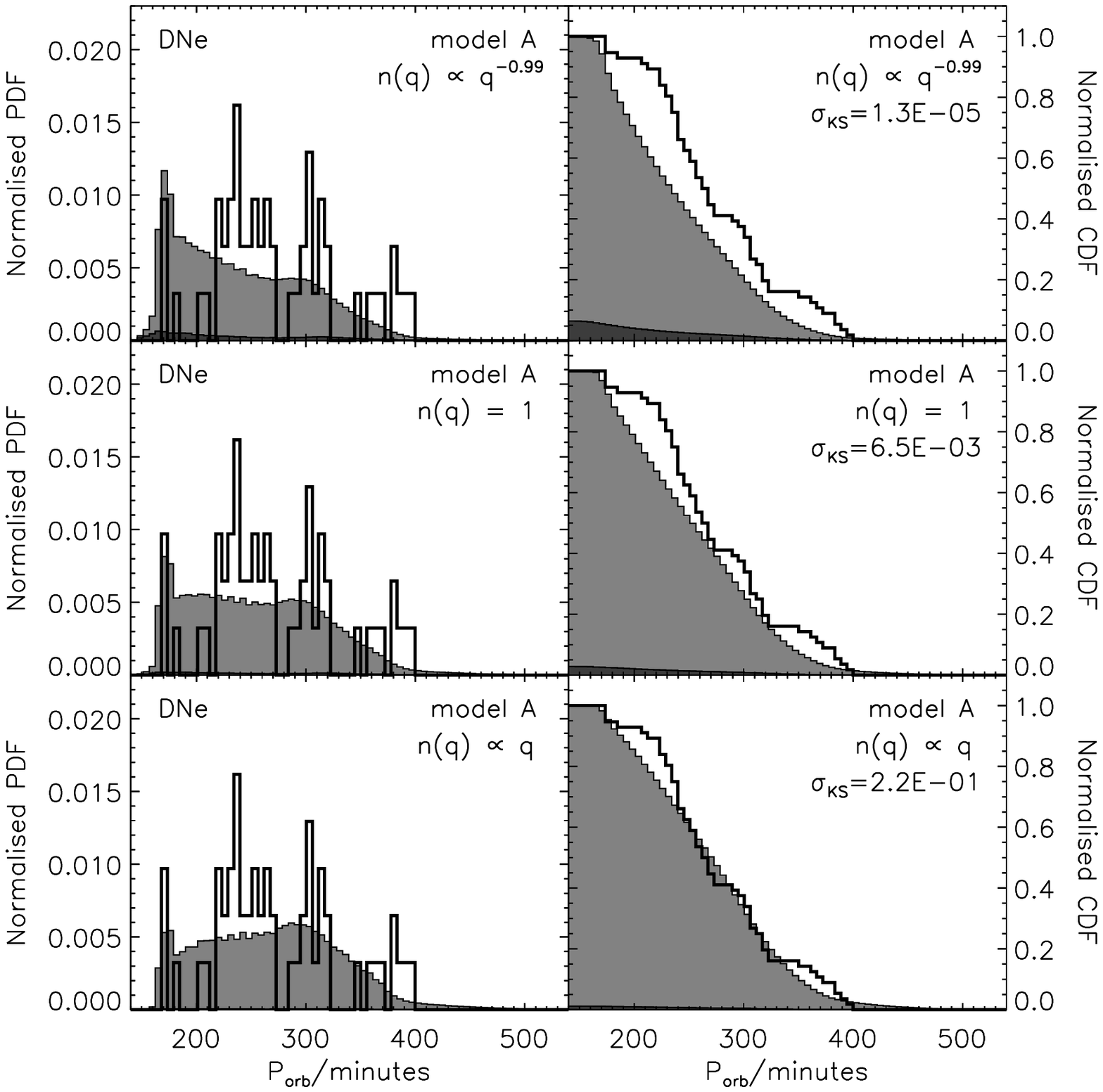}} \hspace{0.2cm}
\resizebox{8.0cm}{!}{\includegraphics{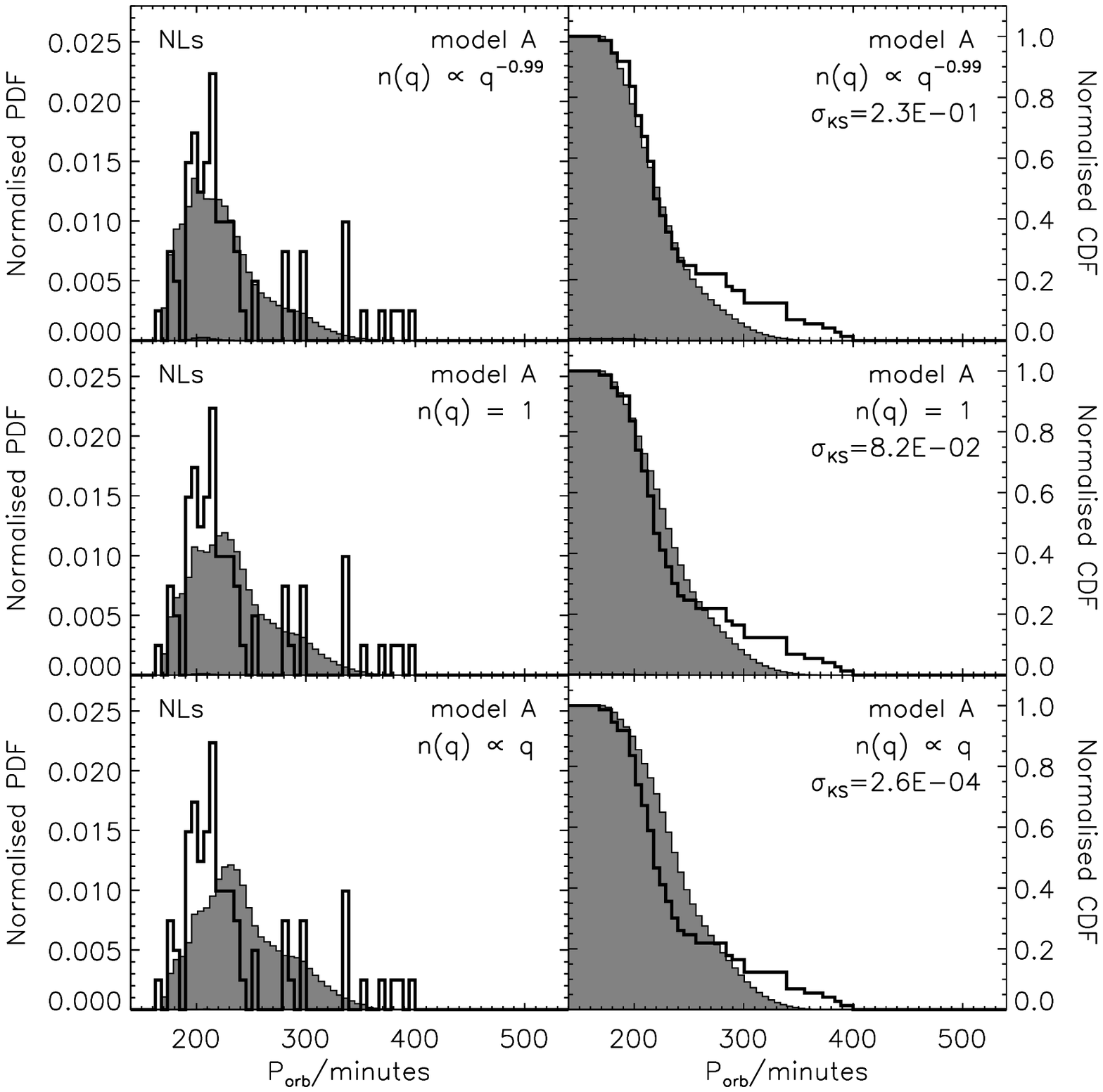}} \hspace{0.2cm}
\caption{Same as Fig.~\ref{pint}, but for a stronger magnetic braking rate with a characteristic time scale of $4 \times 10^9$\,years.}
\label{pintMB2}
\end{figure*}

\begin{figure}
\center
\resizebox{8.0cm}{!}{\includegraphics{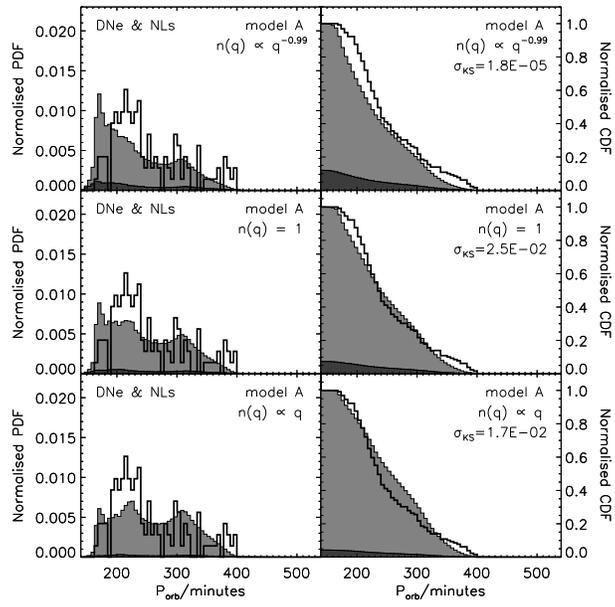}}
\caption{Same as Fig.~\ref{pint0}, but based on the conservative mass transfer stability criterion instead of the non-conservative one.}
\label{pintC}
\end{figure}

Lada (2006) has suggested that the fraction of stars formed in binaries may be substantially lower for stars of spectral type M and later than for stars of earlier spectral types. Such a reduction in the binary fraction can affect the contribution to the orbital period distribution of CVs with donors initially less massive than $0.5\,M_\odot$, which, in our models, spend most of their life time at periods shorter than 4\,hr. Reducing the binary fraction for low-mass stars by a factor of two therefore removes the peak in the theoretical intrinsic DN orbital period distributions at $\sim 170$\,minutes (see Fig.~\ref{pint}), rendering the distributions almost completely flat between 170 and 300\,minutes. The bias corrected distributions change to a much smaller extent since the peak at 170\,minutes is already suppressed for these distributions (see Fig.~\ref{pwgt}). The intrinsic and bias-corrected NL populations are even less affected as NL systems typically have larger minimum periods than DNe.

In conclusion, the main features of the non-magnetic CV orbital period distribution 
can be understood by incorporating the angular momentum loss mechanism associated 
with CB disks in the evolution of CVs.  The simulations reported here reveal that 
systems with main sequence-like donors can undergo a period bounce near the upper 
edge of the observed orbital period gap at 2.75 hr, producing good agreement with 
the orbital period distribution of the combined dwarf novae and nova-like populations. 
As shown in Paper~I, the CB disk induced orbital angular momentum losses also provide an explanation of the period minimum at 80\,minutes and the lack of accumulation of systems at the period minimim in the observed distribution provided that the fractional rate of mass input into the CB disk is decreased from $3 \times 10^{-4}$ to $10^{-5}$. The cause for this difference is unclear, but it may be due to the fact that systems with longer orbital periods have larger  accretion disks and are more likely to emit winds from the outer disk regions residing in shallower parts of the gravitational potential.  Alternatively, the mass loss from the stellar magnetosphere may be more effective for more massive donors.  We note that such mass loss may be greater in a binary system than in single stars since the location of the outer Lagrangian point $L_2$ of the binary system typically lies within the closed field regions of the magnetospheres of single rotating stars (see, e.g., Mestel \& Spruit 1987). In the CB disk picture, a good agreement between the observed and theoretical orbital period distribution near the lower edge of the period gap furthermore requires an increase, by a factor of 100, in the birth rates of systems forming at periods of 2.25 hr (Paper~I).  Although this is inconsistent with a picture in which all unevolved systems formed above the gap undergo a period bounce, it may provide a diagnostic hint for the underlying birth rate distribution (Politano \& Weiler 2006). 

\begin{figure}
\center
\resizebox{8.0cm}{!}{\includegraphics{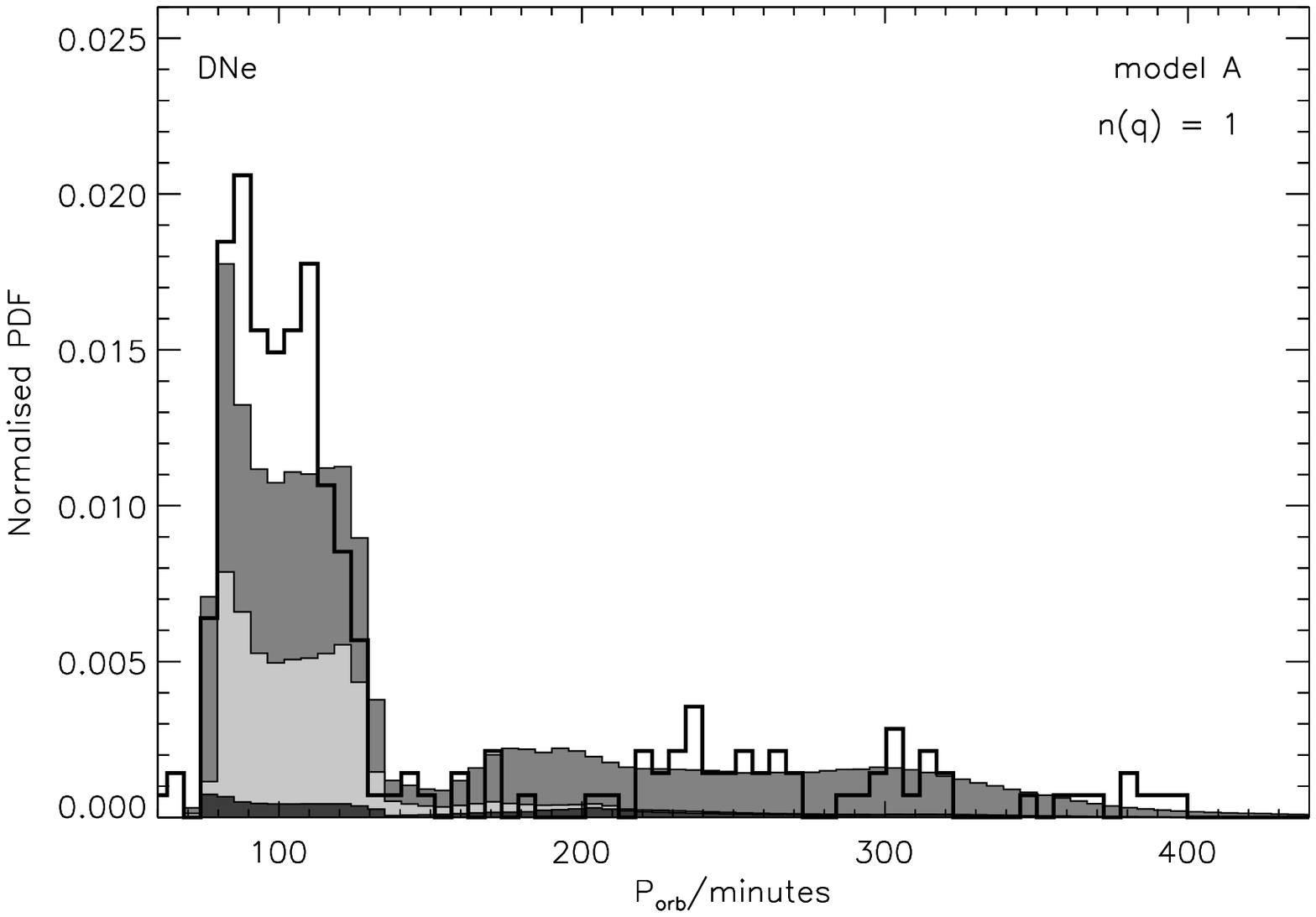}}
\caption{Normalized orbital period distribution of DNe for population synthesis model~A 
and the initial mass ratio distribution $n(q)=1$. As in paper~I, the lower edge of the 
period gap is modeled by increasing the zero-age CV birth rate at 2.25\,hr by a 
factor of 100. Selection effects are taken into account using a detection probability 
factor $W=d\,L_{\rm acc}$. The light, intermediate, and dark gray shading 
indicate the fractions of He, C/O, and O/Ne/Mg WD systems contributing to the 
population, and the thick solid line represents the observed orbital period 
distribution of Galactic DNe.}
\label{ptot}
\end{figure}

For illustration, Fig.~\ref{ptot} shows the complete orbital period distribution 
of DNe below and above 2.75\,hrs obtained by combining the results of Paper~I 
and the results presented in this paper in the case of population synthesis 
model~A and the initial mass ratio distribution $n(q)=1$. As in Paper~I, the 
birth rate of systems at orbital periods of 2.75\,hr is increased by a factor 
of 100 to improve the model near the lower edge of the period gap\footnote{In 
view of the period bounce near the upper edge of the period gap of CVs forming 
at periods longer than 2.75\,hr, the increased birth rate can no longer be 
interpreted as a flow of systems from above to below the gap. Instead, we 
attribute the increase to a modification in the formation space of CVs. For 
this reason, we here applied the increased birth rates to all CVs forming at 
periods  $\sim 2.25$\,hr instead of only to those containing C/O WDs as was 
done in Paper~I}. Selection effects are taken into account by weighting the 
contribution of each system to the orbital period distribution by a factor 
$W=d\,L_{\rm acc}$ which is independent of the duty cycle [see Eqs.~(\ref{d1}) and (\ref{d2})]. 
Apart from an underrepresentation of the number of systems below the gap, 
the theoretical orbital period distribution is in good agreement with the 
observed one. As the relative number of systems below and above the gap depends 
strongly on the adopted observational selection factors 
(an observational selection factor $W=d\,L_{\rm acc}^{1.5}$, for instance, 
significantly increases the theoretical deficit of systems below the gap), 
the insufficient number of systems below the period gap in the theoretical 
distribution could point to 
an essential difference in the selection effects for systems above and below 
the period gap. This is particularly important for the full population 
since the mass transfer rates of the dwarf novae vary by more than one order 
of magnitude for these systems.

Finally, we point out that an understanding of the observational selection effects of the DN and the NL populations is necessary in order to properly compare the observed distribution to our simulated theoretical distributions. A comparison with our intrinsic distributions can fruitfully be undertaken once future surveys complete to within a given volume of space have been attained.

\acknowledgements 
 
We thank M. Politano for a critical reading and useful comments on a preprint 
of this paper.  This research was supported in part by the National Science 
Foundation under grants AST 0200876, a David and Lucille Packard Foundation
Fellowship in Science and Engineering grant, NASA ATP grant
NAG5-13236.  Partial support is also acknowledged from the Theoretical 
Institute for Advanced Research in Astrophysics (TIARA) operated under 
Academia Sinica and the National Science Council Excellence Projects 
program in Taiwan administered through grant number NSC 95-2752-M-007-006-PAE. 
Astronomy Research at the Open University is supported by a PPARC Rolling Grant.

\clearpage

\begin{deluxetable}{lc}
\tablecolumns{2}
\tablecaption{Population synthesis model parameters. 
   \label{models} }
\tablehead{ 
   \colhead{model} &  
   \colhead{$\alpha_{\rm CE}\, \lambda_{\rm CE}$} 
   }
\startdata
A     & 0.5 \\
CE1   & 0.1 \\
CE8   & 2.5 \\
DCE1  & 0.5 for case B RLO \\
      & 2.5 for case C RLO \\
DCE5  & 2.5 when $M_{\rm e} < 2\,M_2$ \\
      & 0.5 when $M_{\rm e} \ge 2\,M_2$ \\
\enddata
\end{deluxetable}

\clearpage

\begin{deluxetable}{lcccccccccc}
\tablecolumns{6}
\tabletypesize{\scriptsize}
\tablecaption{Present-day birth rates of Galactic CVs forming with orbital
  periods longer than 2.75hr. The rates can be converted into approximate local birth rates by dividing them by $5 \times 10^{11}\,{\rm pc^3}$ (see, e.g., Willems \& Kolb 2004). For ease of comparison the last column shows the birth rates of CVs forming with orbital periods shorter than 2.75\,hr, taken from Table~2 of Paper~I\tablenotemark{1}.
\label{br}}
\tablehead{
   \colhead{model} & \colhead{He WD} & \colhead{C/O WD} &
   \colhead{O/Ne/Mg WD} & \colhead{Total} & \colhead{CVs birth rates below 2.75\,hr}
   }
\startdata
\multicolumn{6}{c}{$n(q) \propto q$, $0 < q \le 1$} \\ \hline
A    & $2.9 \times 10^5$ ( 7.1\%) & $3.8 \times 10^6$ (92.4\%) & $1.9 \times 10^4$ (0.5\%)    & $4.1 \times 10^6$ & $6.1 \times 10^5$ \\
CE1  & $4.5 \times 10^5$ (10.3\%) & $3.9 \times 10^6$ (89.7\%) & $2.1 \times 10^2$ ($<$0.1\%) & $4.3 \times 10^6$ & $3.2 \times 10^5$ \\
CE8  & $4.2 \times 10^3$ ( 0.2\%) & $2.0 \times 10^6$ (99.0\%) & $1.6 \times 10^4$ (0.8\%)    & $2.0 \times 10^6$ & $3.9 \times 10^5$ \\
DCE1 & $2.9 \times 10^5$ (13.6\%) & $1.9 \times 10^6$ (85.7\%) & $1.4 \times 10^4$ (0.7\%)    & $2.2 \times 10^6$ & $5.3 \times 10^5$ \\
DCE5 & $2.9 \times 10^5$ ( 9.2\%) & $2.9 \times 10^6$ (90.1\%) & $2.2 \times 10^4$ (0.7\%)    & $3.2 \times 10^6$ & $6.2 \times 10^5$ \\
\hline
\multicolumn{6}{c}{$n(q) = 1$, $0 < q \le 1$} \\ \hline
A    & $5.9 \times 10^5$ ( 8.1\%) & $6.7 \times 10^6$ (90.8\%) & $8.6 \times 10^4$ (1.2\%)    & $7.3 \times 10^6$ & $2.4 \times 10^6$ \\
CE1  & $6.8 \times 10^5$ (12.1\%) & $4.9 \times 10^6$ (87.9\%) & $6.3 \times 10^2$ ($<$0.1\%) & $5.6 \times 10^6$ & $1.0 \times 10^6$ \\
CE8  & $1.0 \times 10^4$ ( 0.3\%) & $3.9 \times 10^6$ (98.0\%) & $7.2 \times 10^4$ (1.8\%)    & $4.0 \times 10^6$ & $1.7 \times 10^6$ \\
DCE1 & $5.9 \times 10^5$ (14.0\%) & $3.6 \times 10^6$ (84.4\%) & $6.5 \times 10^4$ (1.5\%)    & $4.2 \times 10^6$ & $2.0 \times 10^6$ \\
DCE5 & $5.9 \times 10^5$ ( 9.4\%) & $5.6 \times 10^6$ (89.0\%) & $9.6 \times 10^4$ (1.5\%)    & $6.2 \times 10^6$ & $2.4 \times 10^6$ \\
\hline
\multicolumn{6}{c}{$n(q) \propto q^{-0.99}$, $0 < q \le 1$} \\  \hline
A    & $2.4 \times 10^4$ ( 8.0\%) & $2.7 \times 10^5$ (89.1\%) & $8.7 \times 10^3$ (2.9\%)    & $3.0 \times 10^5$ & $2.3 \times 10^5$ \\
CE1  & $2.1 \times 10^4$ (13.2\%) & $1.4 \times 10^5$ (86.8\%) & $3.7 \times 10^1$ ($<$0.1\%) & $1.6 \times 10^5$ & $7.2 \times 10^4$ \\
CE8  & $4.8 \times 10^2$ ( 0.3\%) & $1.8 \times 10^5$ (95.9\%) & $7.1 \times 10^3$ (3.8\%)    & $1.9 \times 10^5$ & $1.8 \times 10^5$ \\
DCE1 & $2.4 \times 10^4$ (12.6\%) & $1.6 \times 10^5$ (83.9\%) & $6.5 \times 10^3$ (3.5\%)    & $1.9 \times 10^5$ & $1.8 \times 10^5$ \\
DCE5 & $2.4 \times 10^4$ ( 8.6\%) & $2.4 \times 10^5$ (88.0\%) & $9.6 \times 10^3$ (3.5\%)    & $2.8 \times 10^5$ & $2.3 \times 10^5$ \\
\hline
\multicolumn{6}{c}{$M_2$ from same IMF as $M_1$} \\  \hline
A    & $9.4 \times 10^5$ (10.7\%) & $7.6 \times 10^6$ (86.6\%) & $2.4 \times 10^5$ (2.8\%)    & $8.8 \times 10^6$ & $1.2 \times 10^7$ \\
CE1  & $8.1 \times 10^5$ (17.1\%) & $3.9 \times 10^6$ (82.9\%) & $5.0 \times 10^2$ ($<$0.1\%) & $4.7 \times 10^6$ & $3.6 \times 10^6$ \\
CE8  & $1.9 \times 10^4$ ( 0.4\%) & $5.1 \times 10^6$ (96.2\%) & $1.8 \times 10^5$ (3.5\%)    & $5.3 \times 10^6$ & $9.6 \times 10^6$ \\
DCE1 & $9.4 \times 10^5$ (16.5\%) & $4.6 \times 10^6$ (80.4\%) & $1.8 \times 10^5$ (3.1\%)    & $5.7 \times 10^6$ & $9.4 \times 10^6$ \\
DCE5 & $9.3 \times 10^5$ (11.1\%) & $7.2 \times 10^6$ (85.7\%) & $2.6 \times 10^5$ (3.1\%)    & $8.4 \times 10^6$ & $1.2 \times 10^7$ \\
\enddata 
\tablecomments{Shown are the present-day birth rates (in units of
  numbers of systems per Gyr) of CVs forming at orbital periods longer than 
  2.75h and their decomposition according to the type of WD in the
  system. The fractions of systems forming with different types of WDs
  are indicated between parentheses.}
\tablenotetext{1}{In Paper~I, we artificially increased the birth rates of systems with C/O WDs forming at 2.25\,hr by a factor of 100 in order to get a sufficiently steep lower edge of the period gap in the theoretical orbital period distributions. This increases the total birth rates of systems forming at periods shorter than 2.75\,hr by less than a factor of 3 compared to the birth rates listed in the last column of the table.}
\end{deluxetable}

\clearpage

\begin{deluxetable}{lccccccccc}
\tablecolumns{10}
\setlength{\tabcolsep}{0.05in}
\tabletypesize{\scriptsize}
\rotate
\tablewidth{605pt}
\tablecaption{Total number of DNe and NLs with orbital periods longer than
  2.75\,hr currently populating the Galactic disk\tablenotemark{1,2}. The numbers can be converted into approximate local space densities by dividing them by $5 \times 10^{11}\,{\rm pc^3}$.
\label{num}}
\tablehead{
   \colhead{} & \multicolumn{4}{c}{DNe} &  & 
   \multicolumn{4}{c}{NLs} \\
   \cline{2-5} \cline{7-10} 
   \colhead{model} & \colhead{He WD} & \colhead{C/O WD} &
   \colhead{O/Ne/Mg WD} & \colhead{Total} &  & \colhead{He WD} &
   \colhead{C/O WD} & \colhead{O/Ne/Mg WD} & \colhead{Total} 
   }
\startdata
\multicolumn{10}{c}{$n(q) \propto q$, $0 < q \le 1$} \\ \hline
A    & $1.6 \times 10^4$ (2.2\%) & $6.9 \times 10^5$ (96.7\%) & $8.2 \times 10^3$ (1.1\%) & $7.2 \times 10^5$ &  & $9.5 \times 10^3$ (3.2\%) & $2.9 \times 10^5$ (96.8\%) & $7.8 \times 10^1$ ($<$0.1\%) & $3.0 \times 10^5$ \\
CE1  & $3.4 \times 10^4$ (5.1\%) & $6.2 \times 10^5$ (94.9\%) & $0.0 \times 10^0$ (0.0\%) & $6.5 \times 10^5$ &  & $1.8 \times 10^4$ (5.2\%) & $3.3 \times 10^5$ (94.8\%) & $0.0 \times 10^0$ (0.0\%) & $3.5 \times 10^5$ \\
CE8  &                          $<$0.1\% & $3.9 \times 10^5$ (98.4\%) & $6.1 \times 10^3$ (1.6\%) & $3.9 \times 10^5$ &  &                           $<$0.1\% & $1.4 \times 10^5$ (99.9\%) & $1.5 \times 10^2$ (0.1\%) & $1.4 \times 10^5$ \\
DCE1 & $1.6 \times 10^4$ (4.0\%) & $3.7 \times 10^5$ (94.4\%) & $6.0 \times 10^3$ (1.5\%) & $3.9 \times 10^5$ &  & $9.5 \times 10^3$ (6.6\%) & $1.3 \times 10^5$ (93.3\%) & $1.4 \times 10^2$ (0.1\%) & $1.4 \times 10^5$ \\
DCE5 & $1.6 \times 10^4$ (2.4\%) & $6.2 \times 10^5$ (96.2\%) & $9.0 \times 10^3$ (1.4\%) & $6.4 \times 10^5$ &  & $9.5 \times 10^3$ (4.7\%) & $1.9 \times 10^5$ (95.2\%) & $9.1 \times 10^1$ ($<$0.1\%) & $2.0 \times 10^5$ \\
\hline
\multicolumn{10}{c}{$n(q) = 1$, $0 < q \le 1$} \\ \hline
A    & $2.9 \times 10^4$ (1.9\%) & $1.5 \times 10^6$ (95.4\%) & $4.2 \times 10^4$ (2.7\%) & $1.5 \times 10^6$ &  & $1.8 \times 10^4$ (3.8\%) & $4.4 \times 10^5$ (96.1\%) & $2.8 \times 10^2$ (0.1\%) & $4.6 \times 10^5$ \\
CE1  & $5.1 \times 10^4$ (5.3\%) & $9.1 \times 10^5$ (94.7\%) & $0.0 \times 10^0$ (0.0\%) & $9.6 \times 10^5$ &  & $2.7 \times 10^4$ (6.6\%) & $3.8 \times 10^5$ (93.4\%) & $0.0 \times 10^0$ (0.0\%) & $4.1 \times 10^5$ \\
CE8  &                           $<$0.1\% & $9.2 \times 10^5$ (96.6\%) & $3.3 \times 10^4$ (3.4\%) & $9.5 \times 10^5$ &  &                         $<$0.1\% & $2.4 \times 10^5$ (99.8\%) & $6.0 \times 10^2$ (0.2\%) & $2.5 \times 10^5$ \\
DCE1 & $2.9 \times 10^4$ (3.2\%) & $8.6 \times 10^5$ (93.4\%) & $3.1 \times 10^4$ (3.4\%) & $9.2 \times 10^5$ &  & $1.8 \times 10^4$ (7.2\%) & $2.3 \times 10^5$ (92.6\%) & $5.5 \times 10^2$ (0.2\%) & $2.4 \times 10^5$ \\
DCE5 & $2.9 \times 10^4$ (2.0\%) & $1.4 \times 10^6$ (94.8\%) & $4.7 \times 10^4$ (3.2\%) & $1.5 \times 10^6$ &  & $1.8 \times 10^4$ (5.1\%) & $3.3 \times 10^5$ (94.8\%) & $3.3 \times 10^2$ (0.1\%) & $3.4 \times 10^5$ \\
\hline
\multicolumn{10}{c}{$n(q) \propto q^{-0.99}$, $0 < q \le 1$} \\ \hline
A    & $ 1.1\times 10^3$ (1.4\%) & $7.1 \times 10^4$ (92.5\%) & $4.7 \times 10^3$ (6.1\%) & $7.7 \times 10^4$ &  & $6.5 \times 10^2$ (4.3\%) & $1.4 \times 10^4$ (95.5\%) & $2.0 \times 10^1$ (0.1\%) & $1.5 \times 10^4$ \\
CE1  & $1.5 \times 10^3$ (5.0\%) & $2.9 \times 10^4$ (95.0\%) & $0.0 \times 10^0$ (0.0\%) & $3.0 \times 10^4$ &  & $7.9 \times 10^2$ (7.8\%) & $9.3 \times 10^3$ (92.2\%) & $0.0 \times 10^0$ (0.0\%) & $1.0 \times 10^4$ \\
CE8  &                           $<$0.1\% & $4.9 \times 10^4$ (93.1\%) & $3.6 \times 10^3$ (6.9\%) & $5.3 \times 10^4$ &  &                          $<$0.1\% & $9.2 \times 10^3$ (99.5\%) & $4.6 \times 10^1$ (0.5\%) & $9.2 \times 10^3$ \\
DCE1 & $1.1 \times 10^3$ (2.2\%) & $4.5 \times 10^4$ (90.7\%) & $3.5 \times 10^3$ (7.1\%) & $5.0 \times 10^4$ &  & $6.5 \times 10^2$ (7.3\%) & $8.2 \times 10^3$ (92.2\%) & $4.2 \times 10^1$ (0.5\%) & $8.9 \times 10^3$ \\
DCE5 & $1.1 \times 10^3$ (1.4\%) & $7.0 \times 10^4$ (91.9\%) & $5.1 \times 10^3$ (6.7\%) & $7.6 \times 10^4$ &  & $6.5 \times 10^2$ (5.3\%) & $1.2 \times 10^4$ (94.6\%) & $2.4 \times 10^1$ (0.2\%) & $1.2 \times 10^4$ \\
\hline
\multicolumn{10}{c}{$M_2$ from same IMF as $M_1$} \\ \hline
A    & $4.2 \times 10^4$ (1.8\%) & $2.2 \times 10^6$ (92.5\%) & $1.4 \times 10^5$ (5.7\%) & $2.4 \times 10^6$ &  & $2.5 \times 10^4$ (6.5\%) & $3.6 \times 10^5$ (93.4\%) & $3.6 \times 10^2$ (0.1\%) & $3.8 \times 10^5$ \\
CE1  & $5.8 \times 10^4$ (6.1\%) & $9.0 \times 10^5$ (93.9\%) & $0.0 \times 10^0$ (0.0\%) & $9.6 \times 10^5$ &  & $3.0 \times 10^4$ (11.2\%) & $2.4 \times 10^5$ (88.8\%) & $0.0 \times 10^0$ (0.0\%) & $2.7 \times 10^5$ \\
CE8  &                          $<$0.1\% & $1.6 \times 10^6$ (93.7\%) & $1.0 \times 10^5$ (6.3\%) & $1.7 \times 10^6$ &  &                             $<$0.1\% & $2.3 \times 10^5$ (99.6\%) & $8.6 \times 10^2$ (0.4\%) & $2.3 \times 10^5$ \\
DCE1 & $4.2 \times 10^4$ (2.7\%) & $1.4 \times 10^6$ (90.9\%) & $1.0 \times 10^5$ (6.5\%) & $1.6 \times 10^6$ &  & $2.5 \times 10^4$ (10.7\%) & $2.0 \times 10^5$ (88.9\%) & $7.9 \times 10^2$ (0.3\%) & $2.3 \times 10^5$ \\
DCE5 & $4.2 \times 10^4$ (1.7\%) & $2.2 \times 10^6$ (92.0\%) & $1.5 \times 10^5$ (6.2\%) & $2.4 \times 10^6$ &  & $2.5 \times 10^4$ (7.5\%) & $3.0 \times 10^5$ (92.4\%) & $4.2 \times 10^2$ (0.1\%) & $3.3 \times 10^5$ \\
\enddata 
\tablecomments{Shown are the total number of DNe and NLs with orbital periods
  longer than 2.75\,hr currently populating the Galactic disk. The
  numbers only reflect systems that were formed above 2.75\,hr and
  thus do not account for systems evolving through the period gap from
  periods below 2.75\,hr. The relative contributions of systems with
  He, C/O, and O/Ne/Mg WDs to the population are indicated between
  parentheses. Approximate space densities can be obtained by dividing
  the absolute numbers of systems by $5 \times 10^{11}\,{\rm pc^3}$
  (see Willems \& Kolb 2004).}
\tablenotetext{1}{There is a slight mismatch between the criterion separating systems with evolved donor stars from systems with unevolved donor stars in the BiSEPS binary population synthesis code used to construct the zero-age CV populations and the full binary evolution code used to construct the CV evolutionary tracks. Because of this, the zero-age CV populations contain some systems with donors initially more massive than $1\,M_\odot$, while the evolved populations do not. This discrepancy leads to the absence of systems with O/Ne/Mg WDs in the present-day population for model CE1 as all O/Ne/Mg WD systems in this model are born with donor stars more massive than $1\,M_\odot$.}
\tablenotetext{2}{For model CE8, the component masses and orbital periods of the zero-age CVs with He WD accretors fall outside the boundaries of the grid of evolutionary tracks. For this model, their contribution to the CV population above 2.75\,hr is therefore not incorporated in the calculations. Based on their relatively small birth rates and faster evolutionary time scales than the more abundant C/O WD systems, we estimate them to contribute less than 0.1\% to the total CV population above the period gap.}
\end{deluxetable}

\clearpage

\begin{deluxetable}{lccccccccc}
\tablecolumns{10}
\setlength{\tabcolsep}{0.05in}
\tabletypesize{\scriptsize}
\tablecaption{Intrinsic fractions of DNe and NLs forming above 2.75h that are
  evolving towards shorter and longer orbital periods. 
\label{prepost}}
\tablehead{
   \colhead{} & \multicolumn{4}{c}{DNe} &  & 
   \multicolumn{4}{c}{NLs} \\
   \cline{2-5} \cline{7-10} 
   \colhead{model} & \colhead{He WD} & \colhead{C/O WD} &
   \colhead{O/Ne/Mg WD} & \colhead{Total} &  & \colhead{He WD} &
   \colhead{C/O WD} & \colhead{O/Ne/Mg WD} & \colhead{Total} 
   }
\startdata
\multicolumn{10}{c}{$n(q) \propto q$, $0 < q \le 1$} \\ \hline
A    & $<$0.1/2.2 & 67.7/28.9 & 0.9/0.3 & 68.6/31.4 & & $<$0.1/3.2 & 47.9/48.9 & $<$0.1/$<$0.1 & 47.9/52.1 \\
CE1  &    0.1/5.1 & 65.4/29.5 & 0.0/0.0 & 65.4/34.6 & & $<$0.1/5.2 & 47.8/47.1 &    0.0/0.0    & 47.8/52.2 \\
CE8  &    0.0/0.0 & 69.1/29.3 & 1.2/0.3 & 70.3/29.7 & &    0.0/0.0 & 48.2/51.7 & $<$0.1/0.1    & 48.2/51.8 \\
DCE1 &    0.1/4.0 & 66.1/28.2 & 1.2/0.3 & 67.4/32.6 & & $<$0.1/6.6 & 45.1/48.2 & $<$0.1/0.1    & 45.1/54.9 \\
DCE5 & $<$0.1/2.4 & 66.4/29.7 & 1.1/0.3 & 67.5/32.5 & & $<$0.1/4.7 & 41.0/54.2 & $<$0.1/$<$0.1 & 41.0/59.0 \\
\hline
\multicolumn{10}{c}{$n(q) = 1$, $0 < q \le 1$} \\ \hline
A    & $<$0.1/1.9 & 62.0/33.4 & 1.9/0.8 & 63.9/36.1 & & $<$0.1/3.8 & 43.4/52.7 & $<$0.1/$<$0.1 & 43.4/56.6 \\
CE1  &    0.1/5.3 & 61.2/33.4 & 0.0/0.0 & 61.3/38.7 & & $<$0.1/6.5 & 43.3/50.2 &    0.0/0.0    & 43.3/56.7 \\
CE8  &    0.0/0.0 & 62.3/34.3 & 2.5/1.0 & 64.8/35.2 & &    0.0/0.0 & 43.7/56.0 &    0.1/0.2    & 43.8/56.2 \\
DCE1 & $<$0.1/3.2 & 60.0/33.3 & 2.4/1.0 & 62.5/37.5 & & $<$0.1/7.2 & 40.6/52.0 &    0.1/0.2    & 40.7/59.3 \\
DCE5 & $<$0.1/2.0 & 60.7/34.2 & 2.2/1.0 & 62.9/37.1 & & $<$0.1/5.1 & 37.3/57.5 & $<$0.1/0.1    & 37.3/62.7 \\
\hline
\multicolumn{10}{c}{$n(q) \propto q^{-0.99}$, $0 < q \le 1$} \\ \hline
A    & $<$0.1/1.4 & 54.2/38.3 & 3.7/2.4 & 57.9/42.1 & & $<$0.1/4.3 & 38.7/56.8 & $<$0.1/0.1 & 38.8/61.2 \\
CE1  &    0.1/5.0 & 56.7/38.2 & 0.0/0.0 & 56.8/43.2 & & $<$0.1/7.8 & 38.6/53.6 &    0.0/0.0 & 38.6/61.4 \\
CE8  &    0.0/0.0 & 53.8/39.4 & 4.4/2.4 & 58.2/41.8 & &    0.0/0.0 & 39.1/60.4 &    0.1/0.4 & 39.3/60.7 \\
DCE1 & $<$0.1/2.2 & 52.1/38.6 & 4.5/2.6 & 56.6/43.4 & & $<$0.1/7.3 & 36.3/56.0 &    0.1/0.3 & 36.4/63.6 \\
DCE5 & $<$0.1/1.4 & 53.1/38.8 & 4.1/2.6 & 57.2/42.8 & & $<$0.1/5.2 & 33.8/60.8 & $<$0.1/0.1 & 33.8/66.2 \\
\hline
\multicolumn{10}{c}{$M_2$ from same IMF as $M_1$} \\ \hline
A    & $<$0.1/1.7 & 48.1/44.4 & 3.0/2.7 & 51.1/48.9 & & $<$0.1/6.5  & 33.1/60.3 & $<$0.1/0.1 & 33.1/66.9 \\
CE1  &    0.1/6.1 & 50.6/43.2 & 0.0/0.0 & 50.7/49.3 & & $<$0.1/11.1 & 32.4/56.5 &    0.0/0.0 & 32.5/67.5 \\
CE8  &    0.0/0.0 & 48.0/45.8 & 3.5/2.7 & 51.5/48.5 & &    0.0/0.0  & 34.3/65.3 &    0.1/0.3 & 34.5/65.5 \\
DCE1 & $<$0.1/2.6 & 46.3/44.6 & 3.6/2.9 & 49.9/50.1 & & $<$0.1/10.7 & 30.7/58.2 &    0.1/0.3 & 30.8/69.2 \\
DCE5 & $<$0.1/1.7 & 47.1/44.9 & 3.3/2.9 & 50.4/49.6 & & $<$0.1/7.5  & 28.6/63.8 & $<$0.1/0.1 & 28.7/71.3 \\
\enddata
\tablecomments{Shown are the intrinsic fractions of CVs forming above
  2.75h that are still evolving toward the period minimum (prebounce
  systems) and that are evolving away from the period minimum
  (postbounce systems), and their decomposition according to the type
  of WD in the system. The fractions are expressed in percent, with
  the first number in each column corresponding to prebounce systems
  and the second to postbounce systems. }
\end{deluxetable}

\clearpage

\clearpage

\clearpage

\clearpage

\clearpage

\clearpage

\clearpage

\clearpage

\clearpage

\clearpage

\clearpage

\clearpage

\clearpage

\clearpage

\clearpage

\clearpage

\clearpage

\begin{thebibliography}{}
\bibitem{An} Andronov, N., Pinsonneault, M., \& Sills, A. 2003, \apj, 582, 358
\bibitem{Bar} Baraffe, I., \& Kolb, U. 2000, \mnras, 318, 354
\bibitem{Ba} Barker, J., \& Kolb, U. 2003, \mnras, 340, 623
\bibitem{Be} Belle, K. E., Sanghi, N., Howell, S. B., Holberg, J. B., \& Williams, P. T. 2004, \aj, 128, 448
\bibitem{dK} de Kool M. 1992, \aap, 261, 188
\bibitem{Do} Donati, J., Forveille, T., Cameron, A., Barnes, J., Delfosse, 
  X., Jardine, M., \& Valenti, J., 2006, Sci, 311, 633
\bibitem{Du1} Dubus, G., Taam, R. E., \& Spruit, H. C. 2002, \apj,
  569, 395 
\bibitem{Du0} Dubus, G., Campbell, R., Kern, B., Taam, R. E., \& Spruit, H.C. 2004, \mnras, 349, 869
\bibitem{Du} D\"unhuber, H. 1994, PhD Thesis, Ludwig-Maximilians-Universit\"at 
  M\"unchen
\bibitem{Ha} Hameury, J. M., Menou, K., Dubus, G., Lasota, J. P., \&
  Hure, J. M. 1998, \mnras, 298, 1048
\bibitem{Ho} Howell, S.B., Nelson, L.A., \& Rappaport, S. 2001, \apj,
  550, 897
\bibitem{Ho2} Howell, S.B., et al., 2006, \apjl, 646, L65
\bibitem{Hu} Hurley, J.R., Tout, C.A., \& Pols, O.R.\ 2002, \mnras, 329, 897 
\bibitem{Iv} Ivanova, N., \& Taam, R. E. 2003, \apj, 599, 516
\bibitem{Ki} King, A. R., Frank, J., Kolb, U., \& Ritter, H. 1995, 
\apj, 444, L37
\bibitem{Kn} Knigge, C.\ 2006, ArXiv Astrophysics e-prints, arXiv:astro-ph/0609671
\bibitem{Ko} Kolb, U. 1996, ASSL Vol.~208: IAU Colloq.~158: 
Cataclysmic Variables and Related Objects, 433
\bibitem{Ko1} Kolb, U., Rappaport, S., Schenker, K., \& Howell, S. 2001, \apj, 563, 958
\bibitem{Ko2} Kolb, U., \& Willems, B. 2005, The Astrophysics of
  Cataclysmic Variables and Related Objects, Eds. J.M. Hameury and
  J.P. Lasota, ASP Conference Series, 330, 17 
\bibitem{Kr} Kroupa, P., Tout, C.A., \& Gilmore, G. 1993, \mnras, 262,
  545
\bibitem{La} Lada, C.~J.\ 2006, \apjl, 640, L63
\bibitem{Li} Li, J. K., Wu, K. W., \& Wickramasinghe, D. T. 1994, \mnras, 268, 61
\bibitem{Ma} Martin, R. G., \& Tout, C. A. 2005, \mnras, 358, 1036
\bibitem{Mc} McDermott, P. N., \& Taam, R. E. 1989, \apj, 342, 1019
\bibitem{Me} Mestel, L., \& Spruit, H. C. 1987, \mnras, 226, 57
\bibitem{Pa0} Patterson, J. 1984, \apjs, 54, 44
\bibitem{Pat} Patterson, J., Kemp, J., Harvey, D. A., Fried, R. E., Rea, R., 
Monard, B., Cook, L. M. et al. 2005, \pasp, 117, 1204
\bibitem{Po} Politano, M. 1996, \apj, 465, 338
\bibitem{Po2} Politano, M., Weiler, K.P. 2006, submitted to ApJ 
\bibitem{Ra} Rappaport, S., Verbunt, F., \& Joss, P. C. 1983, \apj, 275, 713
\bibitem{Ri} Ritter H., Kolb U. 2003, \aap, 404, 301 (update RKcat7.6)
\bibitem{Sh} Shafter, A. W. 1992, \apj, 394, 268
\bibitem{Sha} Shara, M. M., Livio, M., Moffat, A. F. J., \& Orio, 
M. 1986, \apj, 311, 163
\bibitem{Sk} Skumanich, A. 1972, \apj, 171, 565
\bibitem{Sp} Spruit, H. C., \& Ritter, H. 1983, \aap, 124, 267
\bibitem{SpT} Spruit, H. C., \& Taam, R. E. 2001, \apj, 548, 900
\bibitem{Ta} Taam, R. E., Sandquist, E. L., \& Dubus, G. 2003, \apj, 
  592, 1124 
\bibitem{Tu} Tutukov, A., \& Yungelson, L.\ 1979, 
IAU Symp.~ 83: Mass Loss and Evolution of O-Type Stars, 83, 401
\bibitem{Ve} Verbunt, F., \& Zwaan, C. 1981, \aap, 100, L7
\bibitem{WK1} Willems, B., Kolb, U. 2002, \mnras, 337, 1004
\bibitem{WK2} Willems, B., Kolb, U. 2004, \aap, 419, 1057
\bibitem{cbcv} Willems, B., Kolb, U., Sandquist, E.L., Taam, R.E.,
  Dubus, G., 2005, \apj, 635, 1263 (Paper~I)
\end{thebibliography}
\end{document}